\documentclass[review]{elsarticle}
\usepackage{lineno,hyperref}
\modulolinenumbers[5]

\usepackage{amsmath}
\usepackage{amssymb}

\usepackage{graphicx}
\usepackage{subfig}
\usepackage{float}
\usepackage{algorithm,algpseudocode}
\usepackage{color}
\newcommand{\tcr}{\textcolor{blue}}
\newtheorem{rem}{Remark}

\usepackage{url}
\usepackage{breakurl}

\newcommand{\Z}{\mathcal{Z}_1(\rho)}

\begin{document}

\begin{frontmatter}

\title{Resource Allocation and Interference Management in OFDMA-based VLC Networks\tnoteref{mytitlenote}}
\tnotetext[mytitlenote]{This work was supported by the German Research Foundation (DFG) -- FeelMaTyc (329885056).}

\author[mymainaddress]{Marwan Hammouda
\corref{mycorrespondingauthor}}
\cortext[mycorrespondingauthor]{Corresponding author}
\ead{marwan.hammouda@ikt.uni-hannover.de}

\author[mysecondaryaddress]{Anna Maria Vegni}
\author[mythirdaddress]{Harald Haas}
\author[mymainaddress]{J\"{u}rgen Peissig}

\address[mymainaddress]{Institute of Communications Technology, Leibniz Universit\"{a}t Hannover, 
30167 Hannover, Germany. Email: \{marwan.hammouda, and peissig\}@ikt.uni-hannover.de.}
\address[mysecondaryaddress]{Department of Engineering, Roma TRE University, 00146 Roma, Italy. \\Email: annamaria.vegni@uniroma3.it.}
\address[mythirdaddress]{Institute for Digital Communications, Li-Fi Research and Development Centre, the University 
of Edinburgh,  Edinburgh, UK, EH9 3JL. Email: h.haas@ed.ac.uk.}

\begin{abstract}
Resource allocation and interference management are two main issues that need to be carefully addressed in visible light communication (VLC) systems. In this paper, we propose a resource allocation scheme that can also handle the inter-cell interference in a multi-user VLC system that employs orthogonal frequency division multiple access (OFDMA). Particularly, we suggest dividing the cell coverage into two non-overlapping zones and performing a two-step resource allocation process, in which each step corresponds to a different level of allocating resources, i.e., zone and user levels. We initially define both zones in terms of the physical area and the amount of allocated resources and we investigate the impact of illumination requirements on defining both zones. Through simulations, we evaluate the  performance of the proposed scheme in terms of the area spectral efficiency and the fairness level between the two zones. We show that both performance measures can be potentially enhanced by carefully setting a certain design parameter that reflects the priority level of the users located in the region around the cell center, denoted as Zone 0, in terms of the achievable rate. We further eventuate the system performance in a realistic transmission scenario using a simulation tool.
\end{abstract}


\end{frontmatter}

\section{Introduction}
The increasing demand for wireless data has been observed by many studies. Recently, Cisco systems reported $2.5$ exabytes of global mobile data traffic per month at the end of $2014$, 
which is expected to further increase $9$-fold by $2019$~\cite{Cisco}. This rapid increase in the wireless traffic has motivated many researchers to find new technologies that are capable of meeting the required bandwidth, and two potential solutions have been mainly proposed. The first solution is based on a more efficient utilization of the currently used radio frequency (RF) spectrum, which suffers from the underutilization problem since the majority of this spectrum is licensed. In this context, cognitive radio (CR) has been proposed as a solution to improve the spectrum utilization by allowing unlicensed 
users to access the spectrum under some constraints~\cite{haykin2005cognitive}. On the other hand, the second solution suggests expanding the existing spectrum by working in new frequency bands. In this regard, the millimeter wave (mmWave) range  \textit{i.e.}, $[30, 300]$~GHz~\cite{rappaport2013millimeter}, and the 
optical spectra~\cite{pohl2000channel,tsonev2013light,afgani2006visible,elgala2011indoor,kahn1997wireless,komine2004fundamental,
jungnickeleuropean,dimitrov2015principles} are promising candidates due to their wide 
bandwidths that can support high data transmissions. 

Recently, the noticeable advances in white light emitting diodes (LEDs) technology motivated utilizing the visible light spectrum for data transmission along with the white LEDs main function 
of illumination~\cite{afgani2006visible,elgala2011indoor,kahn1997wireless,komine2004fundamental,
jungnickeleuropean,dimitrov2015principles}. In addition to providing the required bandwidth, using white LEDs for data transmission has many advantages over the RF technology. For example, they are cheap, energy efficient and no extra infrastructure is needed since white LEDs would be already installed for lighting. Furthermore, using white LEDs for data transmission 
would inherently ensure data security since light does not propagate through walls. 

To further increase the spectral efficiency of VLC systems, orthogonal frequency division multiplexing (OFDM) modulation scheme is widely 
used~\cite{elganimi2013performance,armstrong2009ofdm, dimitrov2012signal, mesleh2011performance}. Employing the OFDM scheme has another fundamental advantage that the commonly known OFDM multiple access (OFDMA) scheme can be readily applied to serve multiple users simultaneously by dividing bandwidth and power resources among those users. This is indeed a critical requirement in typical indoor places like offices, airports, and hospitals, where multiple users within the VLC cell are expected to be simultaneously active. Consequently, the problem of allocating the available (and limited) bandwidth and power resources among the existing users becomes an issue that should be well handled in order to optimize the system 
performance. This problem has been addressed by many researchers and different \textit{resource allocation (RA)} approaches have been 
investigated~\cite{wilson2011scheduling,bykhovsky2014multiple, mondal2014sinr, le2015resource, Jin_OA, biagi2015last,chowdhury2014dynamic}. For instance, Soltani \textit{et al.} 
in~\cite{wilson2011scheduling} investigate different 
OFDMA scheduling schemes based on the subcarrier clustering, which can significantly reduce the feedback information sent by users. In addition, Bykhovsky and Arnon in \cite{bykhovsky2014multiple} propose an algorithm to manage the interference-constrained subcarrier reuse among different transmitters and power allocation among different subcarriers in a heuristic way. Moreover, Mondal \textit{et al.} in~\cite{mondal2014sinr} develop a joint power and subcarrier allocation scheme to maximize the system throughput under signal-to-interference-and-noise ratio  constraint. In addition, Jin \textit{et al.} in~\cite{Jin_OA} investigate an RA optimization problem applying proportional fairness, while satisfying specific statistical delay constraints. Finally, Chowdhury \textit{et al.}  in~\cite{chowdhury2014dynamic} propose a QoS-based resource allocation scheme in which traffic classes with higher QoS needs are allocated more 
resource than other classes with lower needs. However, the authors assume that the transmission priority is decided by higher layer protocols in the communication system and different users with the same data class are treated the same regardless of their locations. 

In typical indoor scenarios, adjacent access points normally overlap in order to illuminate the entire indoor space. Subsequently, inter-cell interference is a serious factor that can potentially degrade a VLC network performance. This observation was the key motivation behind many studies that extensively explored interference mitigation techniques for VLC 
networks~\cite{ghimire2012self,chen2015fractional,kazemi2016downlink,kazemi2016spectral,
cui2013performance,van2012inter,Chen}. In this context, Ghimire and Haas in~\cite{ghimire2012self} considered an aircraft cabin environment and proposed a self-organising interference management scheme based on the busy burst (BB) protocol, which introduces an additional signaling overhead. Motivated by the known concept of the fractional frequency reuse (FFR) in cellular RF systems, see e.g., \cite{novlan2011analytical,novlan2010comparison}, Chen \textit{et al.} in~\cite{chen2015fractional} applied the FFR concept in a VLC attocell network. The authors demonstrated a significant improvement over a benchmark system with full frequency reuse scheme in terms of the downlink signal-to-interference-plus-noise ratio and the cell edge spectral efficiency. Subsequently, the same approach was examined in downlink cooperative schemes in~\cite{kazemi2016downlink,kazemi2016spectral}.
Further, Huynh \textit{et al.} in~\cite{van2012inter} proposed using two LED sources with different viewing angles in the optical access point and employing soft frequency reuse (SFR) to mitigate inter-cell interference. In addition, Chen \textit{et al.}~\cite{Chen} propose a coalition formation framework for interference management in VLC networks, so that VLC access points are designed to self-organize into cooperation coalitions for interference cancellation in orthogonal time or frequency domain. \tcr{Finally, the authors 
in~\cite{dowhuszko2017achievable,ma2015coordinated,kizilirmak2017centralized} proposed and evaluated different coordinated multi-point (CoMP) transmission scenarios in which the VLC access points coordinate their transmissions to serve
multiple users simultaneously.}

In this paper, we explore the resource allocation and interference mitigation problems of an OFDMA-based VLC network, and we propose a certain geometrical design that is capable of handling both problems. Different than the aforementioned literature studies, we propose a two-step resource allocation algorithm in which each step corresponds to a different level of the resource allocation process. This paper is an extended version of a previous work~\cite{Marwan_ICC17}, where 
we initially introduced the concept of dividing the cell area into two zones with the main goal of addressing illumination and handover needs. Now, in this paper we apply this concept in the context 
of a more general resource allocation framework and extensive simulation results are also shown in order to validate the effectiveness of the proposed approach. To summarize, the main contributions of this paper are listed as follows:
\begin{itemize}
 \item For an OFDMA-based downlink VLC scenario, we propose a two-step resource allocation process. In the fist step, we divide the cell area into two non-overlapping areas and we address the interplay between the physical area of both zones and the number of allocated subcarriers in each zone under the assumption of an equal-power allocation among subcarriers; 
\item We further address the impacts of different system parameters, including the existence of interfering cells and illumination constraints on defining both zones;

\item In the second step, we apply different power allocation schemes among users in each zone, and we investigate the performance of the proposed scheme in terms of the area spectral efficiency and the fairness level between the two zones;
\item We finally consider a realistic indoor scenario and we use a commercial modeling software to display both zones and evaluate the performance of the proposed scheme.
\end{itemize}

\tcr{We remark that  in this paper the proposed framework is not designed only for OFDMA-based systems, but it is also  valid for any other resource-sharing schemes. For instance,  time division multiple access (TDMA) and  code division multiple access (CDMA) schemes can be used by considering the allocated time slots and the codes, respectively, instead of the subcarriers, as described later. The motivation behind the use of OFDMA scheme in this paper is two-fold \textit{i.e.}: (\textit{i})  the OFDMA can be easily integrated with the multi-carrier OFDM modulation scheme, which is widely known and applied to eliminate the channel selectivity in both time and frequency over the individual
subcarriers~\cite{dimitrov2012signal,elgala2009practical}, thus boost the spectral efficiency, and  (\textit{ii})  the OFDMA supports a continuous transmission in time, since all users utilize the entire time span. Thus, illumination requirements can be easily satisfied. For instance, the authors in~\cite{fakidis2013comparison} concluded that OFDMA outperforms  TDMA scheme in typical downlink scenarios, where data communication and illumination are required at the same time.}

The paper is organized as follows. In Section~\ref{sec:system_model}, we provide a general description for VLC networks in indoor environments. 
This description is considered as the pillar for the  proposed approach, which is described in Section~\ref{sec:scheme}. 
In Section~\ref{sec:LOS}, we present a  detailed analysis of the proposed approach, whereas the performance evaluation is provided  in Section~\ref{sec:results}. Finally, conclusions are drawn.
  
\begin{figure}
\centering
\includegraphics[width=0.7\columnwidth]{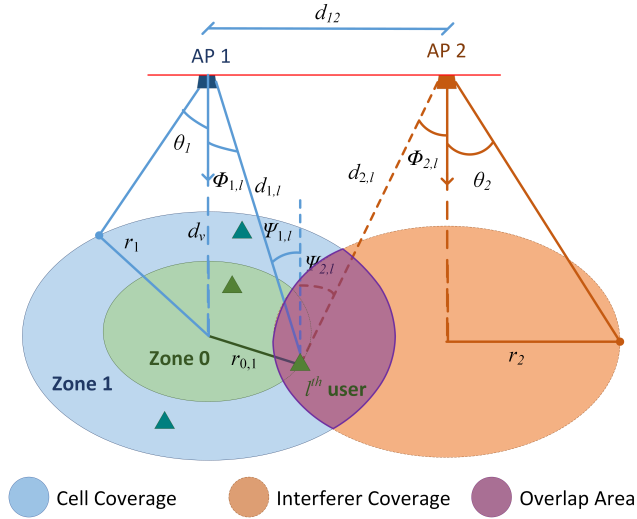}
\caption{Reference scenario with two LED-based APs and multiple users.}
\label{LOS_NLoS}
\end{figure}

\section{System Model}
\label{sec:system_model}
In this paper, we consider a VLC network composed of multiple ``white'' 
 LED-based Access Points (APs). We consider a downlink scenario  and assume that each AP uses OFDMA to serve multiple users within its coverage 
 area. A reference representation of this model is shown in \figurename~\ref{LOS_NLoS} with a number of two APs, \textit{i.e.}, AP1 and AP2. As depicted in \figurename~\ref{LOS_NLoS}, the coverage area of each AP is divided into two regions, \textit{i.e.}, Zone $0$ and Zone $1$. The motivation behind this classification and the definitions of both zones will 
 be detailed later in this paper. For the following analyses, it is sufficient to assume that each AP has a prior knowledge whether each user within its coverage area is located either in Zone $0$ or in Zone $1$, regardless of the user's exact location. The way by which this knowledge is obtained at the APs is beyond the scope of this paper. \tcr{Since the VLC transmission is more suitable for the downlink, we refer to~\cite{shao2014indoor,bao2014protocol}, where the authors propose VLC/RF hybrid scenarios in which VLC is used for the downlink and RF is used for the uplink only. In this context, simple methods can be used in the uplink to provide  the AP with the required information regarding the user's positioning. As for example, the AP can measure the received signal strength and compare it with a pre-defined threshold, such that the user is assumed to be in Zone $0$ if the measured signal is higher than that threshold, whereas the user is assumed to be in Zone $1$ otherwise. Nevertheless, in many cases the location information might be already known by the user for other purposes than data transmission, \textit{e.g.}, indoor positioning and handover management~\cite{biagi2015last,fabini2008location,wisniewski2010location}. In this way, the users can share their locations through the uplink.}
 
Before proceeding, in this paper we mainly consider the following assumptions\footnote{Note that similar assumptions have been considered in many literature studies, see 
\textit{e.g.},~\cite{li2015cooperative,wang2015dynamic}.}: (\textit{i}) all users are connected to the APs via line-of-sight (LoS) links\footnote{While VLC channels are normally composed of both  Line-of-Sight (LoS) and diffuse (multi-path) components, it was observed in~\cite{komine2004fundamental} that, in typical indoor scenarios the majority of the collected energy at the photodiode (more than $95 \%$) comes from the LoS component.}, (\textit{ii}) all APs are directed downwards and all users are directed upwards, and  
(\textit{iii}) each AP produces an ideal cone of light, \textit{i.e.}, its entire light output is projected as a circular lighting field with a hard boundary. In this way, the coverage area of each cell has a circle shape, which is centered at the AP location. 
It is important to remark that the only reason of considering these assumptions in this paper is to make the subsequent analyses tractable towards obtaining closed-form expressions, and hence we can gain deeper insights on the proposed method. However, we emphasize that the main concepts and results in this paper can be applied to general scenarios, in which standard simulation methods, \textit{e.g.}, ray-tracing~\cite{pohl2000channel}, can be used to learn the channel response, run the proposed algorithm, and evaluate the system performance. At this point, 
we refer to Section~\ref{sec:Los_nlos} where we consider a realistic indoor scenario  to assess the performances of our proposed scheme.
 
For the multi-user scenario considered here, let the $l$-th user be served by the $k$-th AP, while being located within the overlapping area between the $k$-th AP and one (or more) adjacent APs. Therefore, the $l$-th user may be subject to inter-cell interference caused by these adjacent cells, which we refer to as \textit{interfering} APs. This scenario is shown in 
\figurename~\ref{LOS_NLoS} for $k=1$ and AP2 being the interfering cell. Since the OFDMA scheme is used by each AP to serve multiple users, then the available bandwidth at each AP is divided into multiple subcarriers, which are then shared among the users served by that cell. From practical perspectives, we assume that each user is equipped with an array of sub-photodetectors, each of which has an analog filter that can be tuned to operate at a different frequency range than the frequency ranges of the other sub-photodetectors in the same array. Indeed, such an array of sub-photodetectors is already available in the market, see \textit{e.g.},~\cite{PDs}. Subsequently, let $B_{k,j}$ and $P_{k,j}$ be, respectively, the bandwidth (in Hz) and the transmitted optical power (in Watts) of the $j$-th subcarrier at the $k$-th AP. Consequently, if that subcarrier is allocated to the $l$-th user laying at a horizontal distance $r$ from the cell center, then the signal-to-interference-and-noise ratio (SINR) can be expressed as
\footnote{This equation is based on the assumption that the neighboring APs \textit{i.e.}, $\{\text{AP}_i\}_{i \neq k}$, transmit different data than that transmitted by the $k$-th AP, and hence they are considered as interference for users served by the $k$-th AP. On the other hand, when all APs transmit, synchronously, the same information, as in broadcast or distributed multiple-input multiple-output scenarios, the signals coming from all APs are considered as useful signals and they accumulate constructively at users served by the $k$-th AP.},\tcr{\footnote{\tcr{We consider a flat-fading channel over the individual subcarriers, which is a reasonable assumption since we only regard the LoS path~\cite{pohl2000channel, 7390429}. Recall that, even if the channel bandwidth is very large and/or the diffuse components are also considered with a frequency selective response, dividing the total bandwidth into subcarriers can totally eliminate the channel selectivity in both time and frequency over the individual subcarriers.}}}
\begin{equation}
\label{SINR_general}
\text{SINR}_{k,l,j}^{(r)} = \frac{(\gamma_l P_{k,j} h_{k,l})^2}{\upsilon N_0 B_{k,j} + \gamma_l^2 \sum_{i \in \mathcal{K}_k}  (P_{i,j} h_{i,l})^2},
\end{equation}
where $h_{k,l}$ is the channel gain between the $l$-th user and the $k$-th AP and $\upsilon$ is the ratio between the average optical power and the average electrical power of the transmitted signal. Herein, we set $\upsilon = 3$, thus the clipping noise is negligible and  the LED can be considered to be working in its linear region~\cite{wang2015dynamic}. Moreover, $\gamma_l$ are $N_0$ are, respectively, the unit-less optical-to-electric conversion efficiency and the noise power spectral density (in A$^{2}$/Hz) of the $l$-th receiver\tcr{\footnote{\tcr{In this paper, we only consider the thermal noise as the main noise source. Nevertheless, we emphasize that the analytical framework and the main results in this study can be directly extended to the general case by considering the signal-dependent noise source.}}}. In \eqref{SINR_general}, the second term of the denominator represents the inter-cell interference induced when the same channel is used for data transmission by the interfering APs, which are  defined by the set $\mathcal{K}_k$, \textit{i.e.,} $\mathcal{K}_1 = \{\text{AP}2\}$ in \figurename~\ref{LOS_NLoS} for $k = 1$, where $P_{i,j}$ is the transmitted power, and $h_{i,l}$ is the channel gain between the $l$-th user and the $i$-th interfering AP. If this interference is mitigated, then (\ref{SINR_general}) reduces to the signal-to-noise ratio (SNR), and we have
\begin{equation}
\label{SNR_general}
\begin{aligned}
\text{SINR}_{k,l,j}^{(r)}  \cong  \text{SNR}_{k,l,j}^{(r)} = \frac{(\gamma_l P_{k,j}h_{k,l})^2}{\upsilon N_0 B_{k,j}}.
\end{aligned}
\end{equation}

Based on the above equations, we consider the achievable rate (\textit{i.e.}, capacity lower bound) of the $l$-th user at distance $r$ from the $k$-th cell center in the 
form~\cite{chaaban2016fundamental,lapidoth2009capacity}
\allowdisplaybreaks
\begingroup
\begin{equation}
\label{rate_r_general}
R_{k,l}^{(r)} = \frac{1}{2} \sum_{j = 1}^{N_s} B_{k,j} \log_2 (1 + c^2 \text{SINR}_{k,l,j}^{(r)} ), \quad \text{[bits/s]}
\end{equation}
\endgroup
for some constant $c$. For instance, $c = \sqrt{e/2 \pi} = 0.93$ if the transmitted light intensity is exponentially distributed~\cite{lapidoth2009capacity}. Without loss of generality, in this paper we set $c = 1$. Above,  $N_s \leq N$ is the number of allocated subcarriers to the $l$-th user and  $N$ is the total number of available subcarriers in each cell. It is reasonable to assume that the cell center of each AP is interference-free. Subsequently, we can immediately observe that the maximum achievable rate over the entire area of the $k$-th AP, denoted as $R_{k}^{\text{max}}$, is achieved when the $l$-th user is located at the cell center (\textit{i.e.}, $r = 0$) 
with all subcarriers being allocated to that user (\textit{i.e.}, 
$N_s = N$). Consequently, we have 
\begin{equation}
\label{max_rate_center}
R_{k}^{\text{max}} = R_{k,l}^{(r=0)} = \sum_{j = 1}^N  B_{k,j} \log_2 (1 + \text{SINR}_{k,l,j}^{(r=0)}).
\end{equation}
 
Throughout this paper, we  assume that all APs have the same optical power ($P_{\text{cell}}$), bandwidth ($B_{\text{cell}}$), and the number of subcarriers ($N$). Also, the total bandwidth in each cell is equally allocated among the subcarriers, \textit{i.e.,} $B_{k,j} = B_{\text{cell}}/N$ for all APs and $j = 1,\dots, N$.

Leveraging on  the aforementioned model, the main research problem addressed in this paper is the RA scheme through which the available bandwidth (subcarriers) and power are efficiently shared among the served users in each cell in order to boost both the system and the per-user throughput, while mitigating the interference terms appeared in (\ref{SINR_general}). In the sequel, we explain our proposed resource allocation scheme to achieve these goals in detail.

\section{Proposed Resource Allocation Scheme}
\label{sec:scheme}
For the system described above, the RA problem is normally formulated as deriving the optimal power and number of subcarriers to be allocated for each user in order to maximize 
either the system or the per-user throughput. Different than that, we propose a geometry-based resource management scheme that can enhance both the system and 
the per-user performance and mitigates the inter-cell interference that might occur in the overlapping areas between adjacent cells. In particular, we propose the following two-step algorithm in which each step corresponds to a different level of RA.

\noindent \textit{\textbf{Step 1: Cell Level --}} \label{Phase_1} In this first step, we aim at improving the system performance on the cell level by enhancing the cell throughput and suppressing the interference induced by the adjacent cells. Herein, we propose achieving these goals by  introducing the concepts of Zone $0$ and Zone $1$, as depicted in \figurename~\ref{LOS_NLoS}, and applying a zone-based RA process. At this point, it is important to remark that these two operations, \textit{i.e.,} defining two zones and applying a zone-level resource allocation, are jointly applied in this step, and hence they are not two separated operations. In detail, Zone $0$ of the $k$-th AP is physically defined as a circle around the cell center with a radius $r_{0,k} \leq r_k$, where $r_k$ is the cell radius. From the RA point of view, in this step we consider an equal power allocation policy among all subcarriers in each cell, and hence we define Zone $0$ in terms of the number of subcarriers allocated for the users located in its area, \textit{i.e.,} $N_{0,k} \leq N$.

Formally, we design Zone $0$ of the $k$-th AP in a way such that if the $l$-th user is placed at the edge of that zone, \textit{i.e.}, at a horizontal distance $r_{0,k}$ from the cell center, and is allocated $N_{0,k}$ subcarriers, then its achieved rate is equal to
\begin{equation}
\label{Zone_0_general_definition}
R_{k,l}^{r=r_{0,k}} = \rho R_{k}^{\text{max}},
\end{equation}
where 
$R_{k}^{\text{max}}$ is given in (\ref{max_rate_center}), and $0 < \rho <1$ is a design parameter. Noting that the achievable rates inside Zone $0$ are higher than those achieved at the zone edge, then the parameter $\rho$ can be seen as the priority measure for the users located in Zone $0$ in terms of the achievable rates. In particular, larger $\rho$ corresponds to a higher priority, whereas smaller $\rho$ implies a lower priority. Subsequently, throughout this paper we denote by $\mathcal{Z}_k(\rho)$ Zone $0$ of the $k$-th AP. Now, by substituting (\ref{rate_r_general}) and (\ref{max_rate_center}) in (\ref{Zone_0_general_definition}), then we can define 
$\mathcal{Z}_k(\rho)$ as 
\begin{equation}
\label{Zone_0_general_definition_1}
\sum_{j = 1}^{N_{0,k}} \log_2(1+ \text{SINR}_{k,l,j}^{(r = r_{0,k})}) = \rho \sum_{j = 1}^N \log_2(1+ \text{SINR}^{(r=0)}_{k,l,j}).
\end{equation}
Recalling that each subcarrier has the same power and bandwidth, then for a given $\rho$ the definition of $\mathcal{Z}_k(\rho)$ should satisfy the following condition:
\begin{equation} 
\label{Zone_0_condition_general}
\begin{aligned}
\mathcal{Z}_k(\rho) =
\bigg \{(r_{0,k},N_{0,k}): \text{SINR}_{k}^{(r = r_{0,k})} \approx ( \text{SINR}^{(r = 0)}_{k})^{\frac{\rho N}{N_{0,k}}} - 1 \bigg \},
\end{aligned}
\end{equation}
where the user and subcarrier indexes have been omitted for notational convenience. Notice that in the above equation we make the approximation  $(1+\text{SINR}^{(r = 0)}_k) \approx \text{SINR}^{(r = 0)}_k$, which is reasonable since the signal strength is expected to be very high at the cell center, \textit{i.e.,} $\text{SINR}^{(r = 0)}_k \gg 1$.
It follows that Zone $1$ in the $k$-th cell, denoted as ${\mathcal{Z}'}_k(\rho)$, is a two-dimensional ring which is defined in terms 
of its width (\textit{i.e.}, $r_{1,k}$), and the number of allocated subcarriers  (\textit{i.e.}, $N_{1,k}$). Formally, we have 
\begin{equation}
{\mathcal{Z}'}_k(\rho) = \{(r_{1,k},N_{1,k}): r_{1,k} = r_k - r_{0,k},N_{1,k} = N - N_{0,k}\},
\end{equation}

Based on the above-described scheme, we highlight the following benefits for dividing each cell into two zones:
\begin{itemize}
\item  Defining different zones and allocating different resources in each zone can be seen as a ``cognitive'' means towards optimizing the coverage area of each AP by carefully selecting the design parameters $\rho$ to satisfy certain system conditions and settings. Later in this paper, we will show the impact of $\rho$ on the system design and performance under different settings and conditions. Nevertheless, the optimization of $\rho$ is beyond the main scope of this paper. Notice that this ``cognitive'' approach is different than the literature studies targeting the VLC coverage area optimization, \textit{e.g.},~\cite{rahaim2013sinr,pergoloni2015coverage} which exploit the physical characteristics of the transmitter and/or receiver, \textit{e.g.}, the viewing angles, to optimize the cell coverage area; 
\item If the overlapping area of any two adjacent cells occurs only over Zone $1$ of both cells, then the inter-cell interference can be mitigated by assigning different frequency bands for the subcarriers allocated in Zone $1$  of the adjacent cells. This process can be realized when each AP shares its information with all adjacent APs, and hence they can perform a collaborative subcarriers assignment, or alternatively, when the VLC network has a central AP that is responsible for the resource allocation in all cells based on a priori knowledge about the entire network\tcr{\footnote{\tcr{Notice that all APs are connected via wires with each other and with the central AP (if exists), and that each of them may have a unique IP address. Therefore, different signaling protocols that are already used in RF wired networks, \textit{i.e.}, in LANs, can be also utilized in our approach.}}}. While the process of assigning the frequency bands in the two zones is not within the scope of this paper, we assume that this process is perfectly performed and hence the inter-cell interference is mitigated, \textit{i.e.}, the coverage area of each cell is interference-free. 
\end{itemize}

\tcr{Following the latter consideration, different reuse methods have been explored in the RF (see \textit{e.g.},~ \cite{novlan2011analytical,novlan2010comparison,
ghaffar2010fractional}) and VLC (see \textit{e.g.},~ \cite{chen2015fractional,van2012inter,ghimire2012self}) literature studies. Any of these techniques can be integrated into our framework and evaluated. Recall that the area and the number of allocated subcarriers in Zone $1$ mainly depend on the new-defined design parameter, $\rho$,for a given number of subcarriers $N$. Thus, the system structure in terms of defining the two zones of adjacent cells can be changed and adapted with respect to the system settings and requirements. It follows that the performance of  the applied reuse method may also depend on the definitions of both zones. Indeed, this idea can be a good seed for a future research study to investigate the interplay between the design parameters, especially $\rho$, and the best  frequency  (subcarriers) reuse (assignment) scheme. }

Notice that this ``cognitive'' method is different than the related resource allocation study~\cite{van2012inter}, where the authors considered different LED viewing angles at each AP to cancel the inter-cell interference. 
Finally, unlike RF networks, typical indoor environments normally have multiple LED-based APs, which are relatively 
close to each other. Therefore, the ability of users to be in Zone $0$ of one of those APs is possible in most of the times, and thus 
defining Zone $0$  with higher $\rho$ values is not expected to impose extra restrictions on the user connectivity or mobility.

\noindent \textit{\textbf{Step 2: User Level --}} \label{Phase_2}
After defining $\mathcal{Z}_k(\rho)$ and $\mathcal{Z}'_k(\rho)$ regions in terms of the radii and the numbers of allocated subcarriers, 
in this step we focus on the resource allocation schemes within each zone. In particular, we aim at sharing the available resources in each zone among  users 
to be served in that zone  with the main target of enhancing the per-user achievable rates. Recalling that the power was assumed to be equally allocated among subcarriers in the first step, the total power allocated in $\mathcal{Z}_k(\rho)$ is $N_{0,k} P_{\text{sub}}$, whereas it is $N_{1,k} P_{\text{sub}}$ in $\mathcal{Z}'_k(\rho)$, where $P_{\text{sub}} = P_{\text{cell}}/N$ is the per-subcarrier transmitted optical power. Different from the first step, both power and subcarrier allocation schemes can be considered in this step and different algorithms can be applied. Note that the number of users that can be served in one zone is restricted by the available sources in that zone. For instance, if $N_{1,k} = 0$ then no users can be supported in Zone $1$, whereas the number of users that can be simultaneously served in Zone $0$ cannot exceed the number of the allocated subcarriers, $N_{0,k}$.

\tcr{From the above descriptions,  in addition to the new-defined parameter $\rho$, considering an OFDMA-based VLC system introduced other main transmission parameters, 
\textit{e.g.}, the total number of subcarriers. We clarify that our main goal is not to provide the optimal values of these parameters that maximize the system performance. Instead, we mainly target  a new approach that can, jointly, handle both the resource allocation and inter-cell interference mitigation problems in an OFDMA-based VLC network. We then explore the impacts of these parameters on the system performance by considering different possible values of these parameters. The natural extension to this paper is to develop certain algorithms to optimize these parameters for given system settings and requirements. However, we keep this step for future studies. }


\section{System Analysis}
\label{sec:LOS}
In this section, we aim at having deeper insights on the proposed resource allocation scheme. In particular, we investigate the impacts of different system parameters on defining Zone $0$, \textit{i.e.} $\mathcal{Z}_k(\rho)$, and Zone $1$, \textit{i.e.}  $\mathcal{Z^{'}}_k(\rho)$, 
as well as  the influence of  illumination constraints on the proposed system. Recall that each APs is directed downwards and each user is directed upwards, then the DC channel gain between the $l$-th user and the $k$-th AP, \textit{i.e.,} $h_{k,l}$, is given by~\cite{barry1993simulation}
\begin{equation}
\label{VLC_gain}
h_{k,l} = \frac{(m_k+1)A_d T(\psi_{k,l}) g(\psi_{k,l}) d_v^{m_k+1}}{2 \pi d_{k,l}^{m_k+3}},
\end{equation}
where $A_d$ is the photodiode  physical area, $d_{k,l}$ is the distance between the $k$-th AP and the $l$-th user, $d_v$ is the vertical distance between the transmitting and receiving planes, and $\psi_{k,l}$ is the angle of incidence with respect to the axis normal to the receiving plane. Furthermore, $m_k = -1/\log_2(\cos(\theta_k))$ is the Lambertian index, where $\theta_k$ is the LED half intensity viewing angle of the $k$-th AP. In (\ref{VLC_gain}), $T(\psi_{k,l})$ is the gain of the optical filter and
\begin{equation}
g(\psi_{k,l}) = \frac{n^2}{\sin^2(\psi_C)}\text{rect}(\psi_{k,l} \leq \psi_C),
\end{equation}
is the optical concentrator gain at the receiver, where $n$ is the refractive index and $\psi_C$ is the field of view  angle of the receiver. Above, $\text{rect}(z)$ is an indicator function such that $\text{rect}(z) = 1$ if $z \leq 1$, and $\text{rect}(z) = 0$ otherwise.

Recall that we assume an equal-power allocation among all subcarriers in the first step (\textit{i.e.}, \textit{Cell Level} in Section~\ref{Phase_1}), with $P_{\text{sub}}$ being the per-subcarrier allocated power in each cell. Now, let the $l$-th user be located at the edge of $\mathcal{Z}_k(\rho)$, \textit{i.e.}, at a horizontal distance $r = r_{0,k}$ from the $k$-th AP center, then the SINR (or simply the signal-to-noise ratio) of the $j$-th subcarrier expressed in (\ref{SINR_general}) can be rewritten as 
\begin{equation}
\label{SINR_general_2}
\begin{aligned}
\text{SINR}_{k,l,j}^{(r = r_{0,k})}  = \frac{[(m_k + 1) \gamma_k P_{\text{sub}} T(\psi_{k,l}) g(\psi_{k,l}) d_v^{m_k+1} A_d]^2}{4 N_0 \upsilon B_{\text{sub}} \pi^2 (r_{0,k}^2 + d_v^2)^{m_k+3}}.
\end{aligned}
\end{equation}
Note that (\ref{SINR_general_2}) follows from the main assumption that $\mathcal{Z}_k(\rho)$ is overlapping-free, and hence interference-free. Nevertheless, such an assumption imposes a certain constraint on defining the area of $\mathcal{Z}_k(\rho)$. In particular, let $d_{kq}$ be the distance between the $k$-th and the $q$-th APs, where $q \in \mathcal{K}_k$ and $\mathcal{K}_k$ is the set of neighboring APs to the $k$-th AP. Consequently, $\mathcal{Z}_k(\rho)$ is overlapping-free only if the following condition holds:
\begin{equation}
\label{eq:r_0_k_inter}
r_{0,k} \leq r_k - \text{max}\{r_k + r_q - d_{kq},0\}_{q \in \mathcal{K}_k} := \Gamma_{k},
\end{equation}
where $r_k$ is the radius of the $k$-th AP. Of course, when the  $k$-th AP is not overlapping with any other  AP, \textit{i.e.,} $\mathcal{K}_k$ is an empty set, then we simply have $\Gamma_{k} = r_k$. In the rest of this section, we focus on a VLC network with two APs, namely AP1 and AP2, as shown in \figurename~\ref{LOS_NLoS}, and we mainly show the analyses with respect to AP1, \textit{i.e.}, 
by setting $k = 1$ in the aforementioned expressions. Nevertheless, the following analyses can be easily extended to more general scenarios. 

\begin{table}[t]
\caption{Parameters used in the simulation results.}
\begin{center}
\begin{tabular}{c|c|c|c|c|c}\hline \hline
{{\bf Parameter}} & {\bf Value} & {{\bf Parameter}} & {\bf Value} & {{\bf Parameter}} & {\bf Value}\\
\hline
$B_{\text{cell}}$ & $20$~MHz &  $P_{\text{cell}} $ &  $9$~W & $d_v$ & $3.5$~m\\
$\psi_C$  & $ 90^{\circ}$ & $A_d$ & $ 1 \,\text{cm}^2$ &  $\gamma$ &  $0.53$ \\
 $n$ & $1.5$ &  $T(\phi_{1,l})$ &$ 1$ &  $N_0$ & $10^{-21}$~$\text{A}^2/$Hz\\
\hline \hline
\end{tabular}
\end{center}
\label{tab_1}
\end{table}%

\subsection{Absence of Illumination Constraints}\label{subsec:no_ill}
We initially assume that no constraints are imposed on the illumination level and  we investigate the influences of different system parameters on defining $\mathcal{Z}_1(\rho)$. \tcr{Notice that this assumption can be applied in some indoor scenarios, such as airports, where the number of users to be served is expected to be larger. Thus, the resource allocation strategy might have a higher priority than the illumination constraints.} In such a case, we can evaluate $r_{0,1}$ for a given $\rho$ from (\ref{eq:r_0_k_inter}) and 
by substituting (\ref{SINR_general_2}) into (\ref{Zone_0_condition_general}), such as 
\begin{equation}
\label{d_1_LOS_N_1}
r_{0,1} = \min \bigg \{\sqrt{\bigg ( \frac{\lambda_1}{(1+\lambda_1 d_v^{-2m_1-6})^{\frac{\rho N}{N_{0,1}}}- 1}\bigg)^{\frac{2}{2m_1+6}} - d_v^2},\Gamma_{1} \bigg \},
\end{equation}
where 
\begin{equation}
\lambda_{1} = \frac{[(m_1 + 1) \gamma_l P_{\text{sub}} T(\psi_{1,l}) g(\psi_{1,l}) d_v^{m_1+1} A_d]^2}{4 N_0 \upsilon B_{\text{sub}} \pi^2}. \label{lambda_1}
\end{equation}

\begin{figure}
\centering
\includegraphics[width=0.8\textwidth]{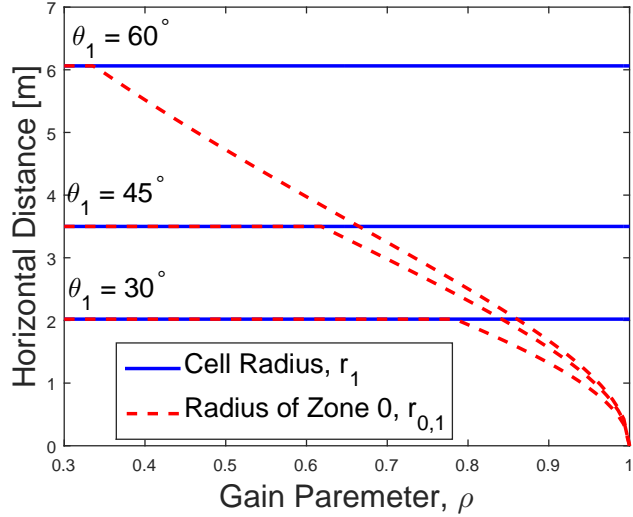}
\caption{Radius of $\mathcal{Z}_1(\rho)$ as a function of $\rho$ for different values of the angle $\theta_1$. Herein, $N = 1$.}
\label{LOS_rho_different_theta}
\end{figure}

Notice that for a given $r_{0,1}$, the number of subcarriers $N_{0,1}$ can be obtained by re-arranging \eqref{d_1_LOS_N_1}, such that 
\begin{equation}
\label{eq:Ns}
N_{0,1} = \bigg \lfloor \frac{\rho N \ln(1+\lambda_1 d_v^{-2m_1-6})}{\ln(1+\lambda_1 (d_v^2 + r_{0,1}^2)^{-m_1-3})}\bigg \rfloor,
\end{equation}
where $\lfloor \cdot \rfloor$ is the floor operator since $N_{0,1} \in \mathbb{N}^{+}$. We initially consider the case when the distance between the two APs, $d_{12}$, is large enough to assume that the two APs are not overlapping, \textit{i.e.,} $\Gamma_{1} = r_1$. In \figurename~\ref{LOS_rho_different_theta}, we show $r_{0,1}$ (\textit{dashed red lines}) as a function of $\rho$ for different values of the LED 
half-intensity angle $\theta_1$, and assuming the parameters in Table~\ref{tab_1}. Herein, we set the number of subcarriers to $N = 1$, and hence we have $N_{0,1} = N$, $P_{\text{sub}}  = P_{\text{cell}}$, and $B_{\text{sub}} = B_{\text{cell}}$. We further show the cell radius  $r_1$ in each case (\textit{blue lines}).  
As we can see, increasing $\rho$ reduces the radius of $\mathcal{Z}_1(\rho)$, and then  we obtain $\rho = 1$ only when $r_{0,1} = 0$, 
\textit{i.e.}, at the cell center. We also note that the angle $\theta_1$ affects $r_{0,1}$ with more effect at lower values of $\rho$.
For instance, the user can achieve up to $80\%$ of the maximum rate when being located at the cell edge for $\theta_1 = 30^{\circ}$, whereas this ratio is reduced to around $40\%$ by increasing the angle to $\theta_1 = 60^{\circ}$. This is expected since reducing the LED angle means that the power is focused in a more confined area, thus higher rates are expected over the cell area.

\begin{figure}
\centering
\subfloat[$\theta = 30^{\circ}$]{\includegraphics[width=0.55\textwidth]{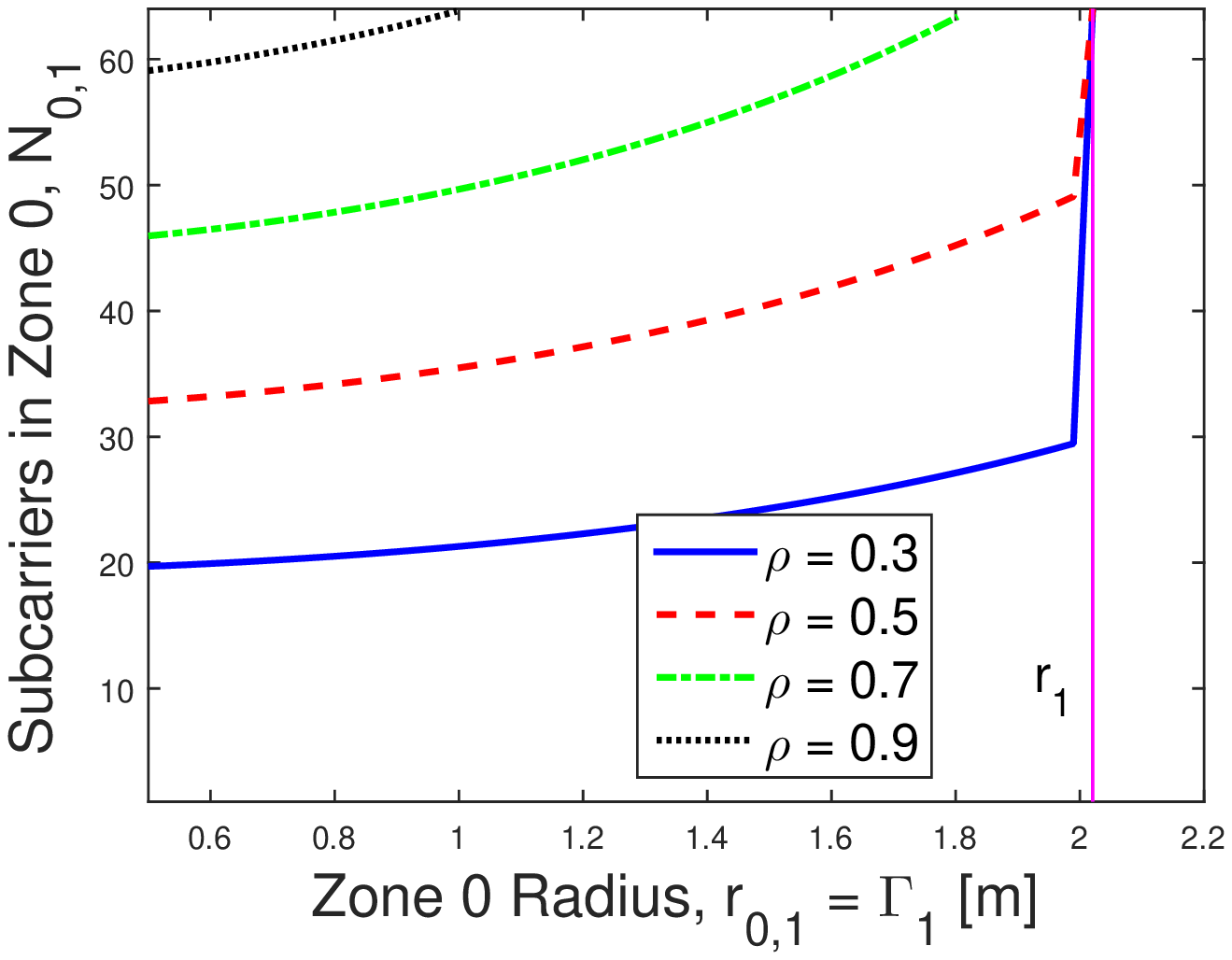}\label{LOS_Ns_different_rho_theta_30}}
\subfloat[$\theta = 60^{\circ}$]{\includegraphics[width=0.55\textwidth]{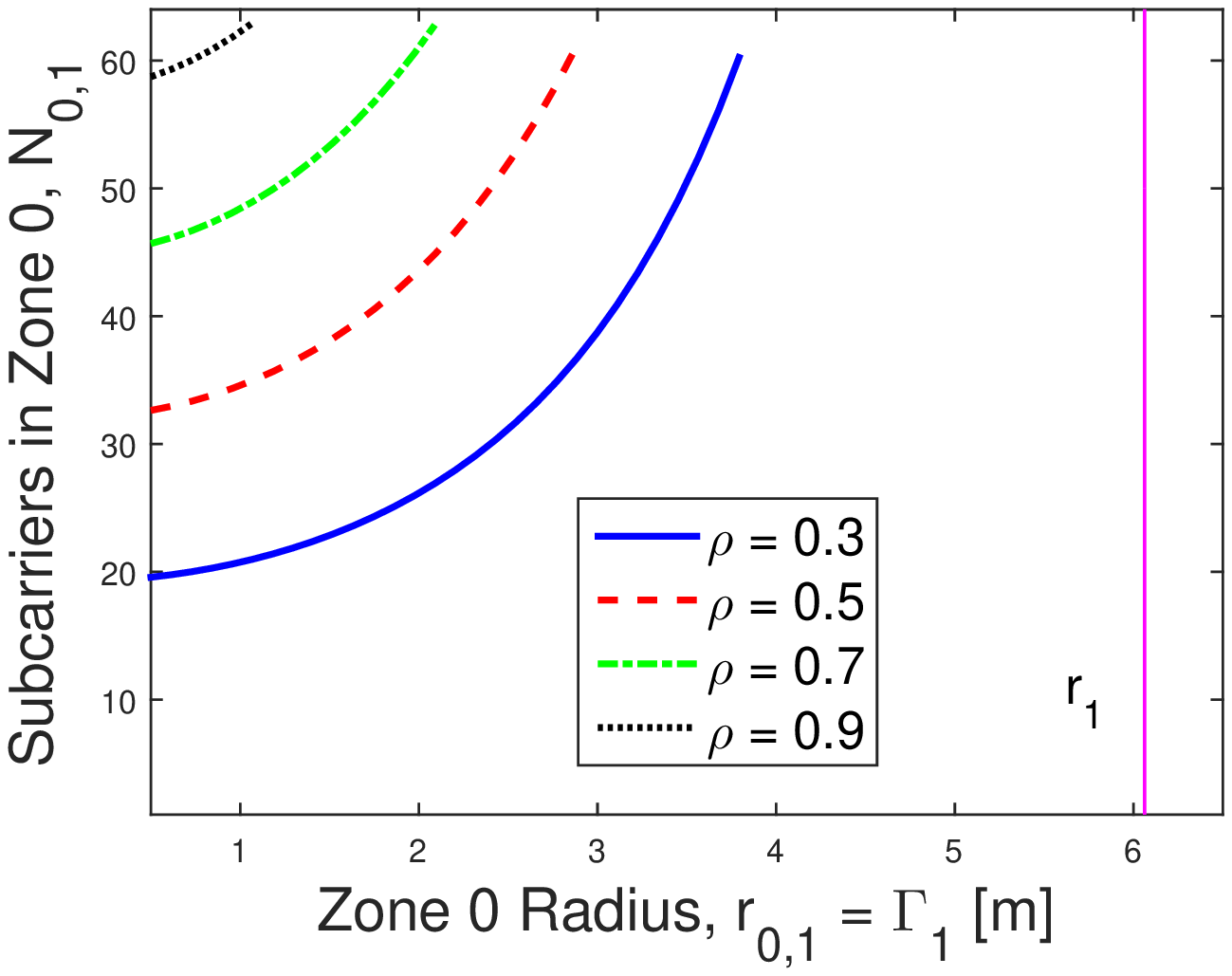}\label{LOS_Ns_different_rho_theta_60}}   
\caption{Number of subcarriers $N_{0,1}$ vs. the radius  $r_{0,1}$ for different values of $\rho$ and $\theta_1$ and $N = 64.$}
\label{LOS_Ns_different_rho_theta}
\end{figure}

For $N = 64$, \figurename~\ref{LOS_Ns_different_rho_theta} depicts the number of allocated subcarriers $N_{0,1}$ as a function of $r_{0,1}$ and for different values of $\rho$ and $\theta = \theta_1 = \theta_2$. In this figure, we vary the value of the distance between the two APs within the range $d_{12} \in [r_1 + 0.5, 2r_1]$, or equivalently $\Gamma_1 \in [0.5,r_1]$. For each value of $d_{12}$, we set the radius of $\mathcal{Z}(\rho)$ to $r_{0,1} = \Gamma_1$ and then we calculate the required number of subcarriers in Zone 0, \textit{i.e.,} $N_{0,1}$ to achieve the considered value of $\rho$. Note that for some values of $\rho$, we might have $N_{0,1} < N$ when $r_{0,1} = r_1$. In such a case, we immediately set $N_{0,1} = N$ to make sure that all resources are used for serving users. As expected, we can clearly observe that increasing the distance $d_{12}$ enables having larger areas of $\mathcal{Z}_1(\rho)$, which further implies that more subcarriers should be allocated to maintain the same value of $\rho$. On the other hand,  when increasing the radius $r_{0,1}$ while keeping the number of subcarriers fixed, the value of $\rho$ decreases, and hence the system performance is expected to degrade. For instance, in \figurename~\ref{LOS_Ns_different_rho_theta} ({b}) for $N_{0,1} = 50$, we have $r_{0,1} = [1.19, 2.4, 3.48]$~m for $\rho = [0.7, 0.5, 0.3]$, respectively. Notice that the radius of $\Z$ can be linked with the number of users located in Zone $0$ by the user density, \textit{i.e.}, the number of users by unit area. Indeed, the user density  in $\Z$ can be expressed as $\mu = N_{pu}/\pi r_{0,1}^2,$ where  $N_{pu}$ is the number of users  placed in Zone $0$. \figurename~\ref{LOS_Ns_different_rho_theta} reveals that, if $N_{pu}$ increases, we should increase both the radius and the allocated subcarriers of $\Z$ correspondingly, in order to maintain the same values of $\rho$ and $\mu$.

Finally, notice that for higher values of $\rho$ (\textit{e.g.}, $\rho = 0.9$ in \figurename~\ref{LOS_Ns_different_rho_theta_30}) and/or a larger cell area (\textit{e.g.}, 
for $\theta_1 = 60^{\circ}$ as in \figurename~\ref{LOS_Ns_different_rho_theta_60}), Zone 0 has a remarkably smaller area that the cell area. On the other hand, Zone $0$ can have the same area as the cell area, with all or some of the subcarriers being allocated, for smaller values of $\rho$ and cell areas. This observation provides an initial insight on the system settings at which the proposed resource allocation scheme based on defining Zone $0$ is expected to have a potential improvement in terms of the system achievable rates. 

From the above results, we can obviously see that the characterization of $\mathcal{Z}_1(\rho)$ highly depends on the parameters $\rho$, $r_{0,1}$ and $N_{0,1}$, which  provide a high flexibility in defining Zone $0$ for each AP based on the considered scenario and system settings. In this paper, we consider the strategy of finding the maximum possible radius $r_{0,1}$ for a given $\rho$ such that $N_{0,1} \leq N$.
We again remark that finding the optimal values of these parameters is not the scope of this paper, and remains an open research 
question to be further investigated in future studies.

\subsection{Effect of Illumination Requirements}\label{subsec:ill}
Recalling that the LED-based APs are originally employed for illumination purposes, ensuring a sufficient amount of light 
over the receiving plane is a crucial requirement. In this regard, the illuminance is the most significant parameter, 
and it characterizes the brightness factor of the illuminated surface. Particularly, let us assume the illuminated surface be located at a horizontal distance $r_{0,1}$ from the AP1 center then the horizontal illuminance  can be calculated as~\cite{grubor2008broadband}
\begin{equation}
\label{illumination_1}
E = I_0 \frac{d_v^{m_1 + 1}}{(r_{0,1}^2 + d_v^2)^{\frac{m_1 + 3}{2}}}, \quad {\rm [lx]}
\end{equation}
where $I_0$ is the maximal luminous intensity [cd]. 
According to the European lighting standard~\cite{Standard}, different brightness levels are required in indoor scenarios depending on specific activities. In this paper, we aim for a brightness span of $[E_{\text{min}},E_{\text{max}}]$ lx within $\mathcal{Z}_1(\rho)$, such that the brightness level at the zone edge fulfills the minimum level $E_{\text{min}}$~[lx], while the illumination level at the cell center does not exceed $E_{\text{max}}$~[lx]. Consequently, by solving \eqref{illumination_1} the radius of $\mathcal{Z}_1(\rho)$ is  limited  as:
\begin{equation}
\label{illumination_2}
r_{0,1} \leq \bigg[\bigg(\frac{I_0 d_v^{m_1+1}}{E_{\text{min}}}\bigg)^{\frac{2}{m_1 + 3}} - d_v^2\bigg]^{\frac{1}{2}}.
\end{equation}
On the other hand, we have the following constraint on the maximal luminous intensity of the light source:
\begin{equation}
E_{\text{min}} d_v^2 \leq I_0 \leq E_{\text{max}} d_v^2.
\end{equation}
Note that, if the illumination level at the cell center equals $E_{\text{max}}$, then $I_0 = E_{\text{max}} d_v^2$, 
and the limit on $r_{0,1}$ in \eqref{illumination_2} can be re-expressed as\cite{Marwan_ICC17}  
\begin{equation}
\label{illumination_3}
r_{0,1} \leq d_v \bigg[\bigg(\frac{E_{\text{max}}}{E_{\text{min}}}\bigg)^{\frac{2}{m_1 + 3}} - 1\bigg]^{\frac{1}{2}} := \Lambda_{1},
\end{equation}
which depends only on the ratio between the minimum and the maximum illumination span $\frac{E_{\text{max}}}{E_{\text{min}}}$, rather than the exact values. To get a better understanding about the effect of fulfilling certain illumination requirements on defining $\mathcal{Z}_1(\rho)$, we assume that the brightness span is defined by $[E_{\text{min}},E_{\text{max}}] = [200,800]$ lx, 
as considered in~\cite{grubor2008broadband}.  In such a case, we can rewrite \eqref{illumination_3} as
\begin{equation}
\label{illumination_4}
r_{0,1} \leq d_v  \bigg(16^{\frac{1}{m_1 + 3}} - 1\bigg)^{\frac{1}{2}},
\end{equation}
which is a function of only the vertical distance $d_v$, and the LED half intensity viewing angle $\theta_1$ through $m_1$. Now by combining (\ref{illumination_3}) and \eqref{d_1_LOS_N_1}, we have the following modified constraint on defining $\mathcal{Z}(\rho)$:
\begin{equation}
\label{eq:radius_final}
r_{0,1} = \min \bigg \{\sqrt{\bigg ( \frac{\lambda_1}{(1+\lambda_1 d_v^{-2m_1-6})^{\frac{\rho N}{N_{0,1}}}- 1}\bigg)^{\frac{2}{2m_1+6}} - d_v^2},\Gamma_{1}, \Lambda_1 \bigg \},
\end{equation}

By comparing the above equation with \eqref{d_1_LOS_N_1}, we can immediately observe the effect of 
considering the illumination requirements on defining the radius 
of $\mathcal{Z}_1(\rho)$. For instance, for $d_v = 3$ m and $\theta_1 = 60^{\circ}$, we have $r_{0,1} \leq d_v \tan(\theta_1) \approx 5.2$~m 
when no illumination requirements are considered, whereas we have $r_{0,1} \leq 3$~m according to \eqref{illumination_4}.

\tcr{
\begin{rem}
The signal processing techniques at the transmitter may have an impact on illumination. As an instance, the authors in~\cite{zhang2016energy} defined the minimum required optical power to satisfy certain illumination requirements when an asymmetrically clipped optical OFDM (ACO-OFDM) technique is employed. However, in this paper we skip the signal processing impact on illumination, and we rely mainly on the intensity profile of the transmitter by defining the received illuminance with respect to the luminous intensity, as expressed in \eqref{illumination_1}. A similar approach was also considered in~\cite{stefan2013area}.
The reason behind following this approach  is twofold. First, recall that allocation framework proposed in this paper can be easily extended to any modulation and resource sharing schemes, thus we provide a generic illumination analysis in the sense that it is independent of the modulation and sharing methods. In this context, we remark that the relation between the photometric and the radiometric quantities, i.e., luminous intensity and power, respectively, is already defined and exits in many studies, see \textit{e.g.},~\cite{rahaim2013sinr,González2014,o2005short} and references therein. Subsequently, the approach provided in this paper can also be easily extended to regard the signal processing part. 
Second, in this work we provide a geometry-based resource allocation scheme, hence it is more reasonable to explore the geometry-based impact of illumination requirements, through the horizontal and vertical distances, i.e., $r_{0,1}$ and $d_v$, respectively. 
\end{rem}
}

\section{Performance Evaluation}\label{sec:results}
So far, we have investigated the effects of different system parameters on defining  $\mathcal{Z}_1(\rho)$ and
$\mathcal{Z}_1^{'}(\rho)$ in terms of the related radius  and the number of allocated subcarriers. In particular, we have the following definitions from the preceding analyses:  (\textit{i}) the radii in $\mathcal{Z}_1(\rho)$ and 
$\mathcal{Z}_1^{'}(\rho)$ are $r_{0,1}$ and $r_{1,1}$, respectively; (\textit{ii}) the number of subcarriers 
in $\mathcal{Z}_1(\rho)$ and $\mathcal{Z}_1^{'}(\rho)$ are $N_{0,1}$ and $N_{1,1}$, respectively. Since each subcarrier was assumed to have a fixed power of $P_{\text{sub}}$, then the total powers allocated 
in $\mathcal{Z}_1(\rho)$ and $\mathcal{Z}_1^{'}(\rho)$ can be calculated, respectively, as $P_{\text{sub}} N_{0,1}$ and $P_{\text{sub}}  N_{1,1}.$

In this section, we study the system performance when considering both zones. We mainly  provide simulation results expressed in terms of area spectral efficiency and the proposed scheme fairness.  For simulation purposes, we assume that each user, either being in Zone $0$ or in Zone $1$, is allocated a single subcarrier. Thus, we set the number of users located in Zone 0 to $N_{pu} = N_{0,1}$ and the number of users located in Zone $1$ to $N_{su} = N_{1,1}$. We further assume that the users located in each zone (either in Zone $0$ or Zone $1$) are uniformly distributed in that zone.

\begin{figure}
\begin{center}
\begingroup\captionof{algorithm}{}\label{algorithm_1}
	\begin{algorithmic}[1] 
    \State {\bf Inputs}: $N$, $B_{\text{sub}}$, $P_{\text{sub}}$, $r_1$, $\Gamma_1$, $\Lambda_1$, $\beta$ and $\rho$;
    \State {\bf Phase I: Cell-Level Resource Allocation}
    \State Set $r_{0,1} = r_{\text{lim}}$, where $r_{\text{lim}}$ is the solution of (\ref{eq:radius_final}) are replacing $N_{0,1}$ with $\beta N$;
    \State Calculate $N_{0,1}$ using (\ref{eq:Ns});\label{step:d_1_Ns}
    \If {$N_{0,1} <= \beta N$}
    \State Set $N_{0,1}^{*} = N_{0,1}$ and $r_{0,1}^{*} = r_{0,1}$;
    \Else
    \State Update $r_{0,1} = r_{0,1} - 0.001$ and return to Step~\ref{step:d_1_Ns};
    \EndIf
    \State Set $\Z = (r_{0,1}^{*},N_{0,1}^{*})$ and $\mathcal{Z}'_1(\rho) = (r_{1,1}^{*},N_{1,1}^{*})$, where $r_{1,1}^{*} = r_1 - r_{0,1}^{*}$ and $N_{1,1}^{*} = N - r_{0,1}^{*}$;
     \State {\bf Phase II: User-Level Resource Allocation}
     \State Given $\Z$ and $\mathcal{Z}_1^{'}(\rho)$ from Phase I;
     \State Set the number of users served by $\Z$ to $N_{pu} = N_{0,1}^{*}$ and by $\mathcal{Z}_1^{'}(\rho)$ to $N_{su} =  N_{1,1}^{*}$;
     \State Locate the users of each zone uniformly in that zone \label{step:locate}; 
    \State Perform power allocation among the users in each zone, e.g., through (\textit{i}) equal power, 
   (\textit{ii}) water-filling, or (\textit{iii}) channel inversion algorithms
   \label{step:power_allocation_0};
      \State  Repeat Steps \ref{step:locate} to \ref{step:power_allocation_0} for the benchmark scenario of $N$ users in the entire cell of radius $r_1$;
     \State Calculate $\eta$ from (\ref{eta})~\label{step:eta};
     \State Repeat Steps \ref{step:locate} to \ref{step:eta} for 10000 times and calculate the average $\eta$.
	\end{algorithmic}
\endgroup
\label{ALGORITHM}
\end{center}
\end{figure}

\tcr{\subsection{Area Spectral Efficiency}}
We initially analyze the system performance from the cell perspective. To that end, we employ the area spectral efficiency (ASE) as the main performance measure. 
ASE was initially introduced in~\cite{alouini1999area} to quantify the spectral efficiency of cellular systems, and is defines the sum achievable rates per unit area supported within a cell. Denoting the ASE of AP$_1$ as $A_{e,1}$ $[\text{bits}/\text{s}/\text{Hz}/\text{m}^2]$, then we have
\begin{equation}
\label{ASE_proposed}
A_{e,1} = \frac{\sum_{t = 1}^{N_{pu}} R_{1,t}^{(r \leq r_{0,1})} + \sum_{j = u}^{N_{su}} R_{1,u}^{(r_{0,1} \leq r < r_{1})}}{ \pi B_{\text{cell}} r_1^2},
\end{equation}
where $R_{1,t}^{(r \leq r_{0,1})}$ and $R_{1,u}^{(r_{0,1} \leq r < r_{1})}$ are, respectively, the data rates of the $t$-th user located in $\Z$ and the $u$-th user located in $\mathcal{Z}_1^{'}(\rho)$, as expressed in (\ref{rate_r_general}) for $N_s = 1$ for both users. For a comparison purpose, we  consider the most commonly used resource allocation approach 
in which all users have the same priority regardless of their locations, \textit{i.e.}, no zones are considered, as the benchmark scenario.
For such a scenario, we denote the ASE as $A_{e,1}^{\text{uni}}$, \textit{i.e.}, 
\begin{equation}
\label{ASE_uniform}
A_{e,1}^{\text{uni}} = \frac{\sum_{q = 1}^{N_{pu}} R_{1,t}^{(r \leq r_{1})}}{ \pi B_{\text{cell}} r_1^2}.
\end{equation}
From (\ref{ASE_proposed}) and (\ref{ASE_uniform}), we define $\eta$ as the performance gain achieved by employing the proposed allocation scheme over the benchmark scheme, \textit{i.e.}, 
\begin{equation}
\label{eta}
\eta = \frac{A_{e,1}}{A_{e,1}^{\text{uni}}} = \frac{\sum_{t = 1}^{N_{pu}} R_{1,t}^{(r \leq r_{0,1})} + \sum_{j = u}^{N_{su}} R_{1,u}^{(r_{0,1} \leq r < r_{1})}}{\sum_{q = 1}^{N_{pu}} R_{1,t}^{(r \leq r_{1})}}.
\end{equation}

\begin{figure}{c}
\centering
\subfloat[Solid (dashed) lines for the present (absent) of illumination constraints.]{\includegraphics[width=0.65\textwidth]{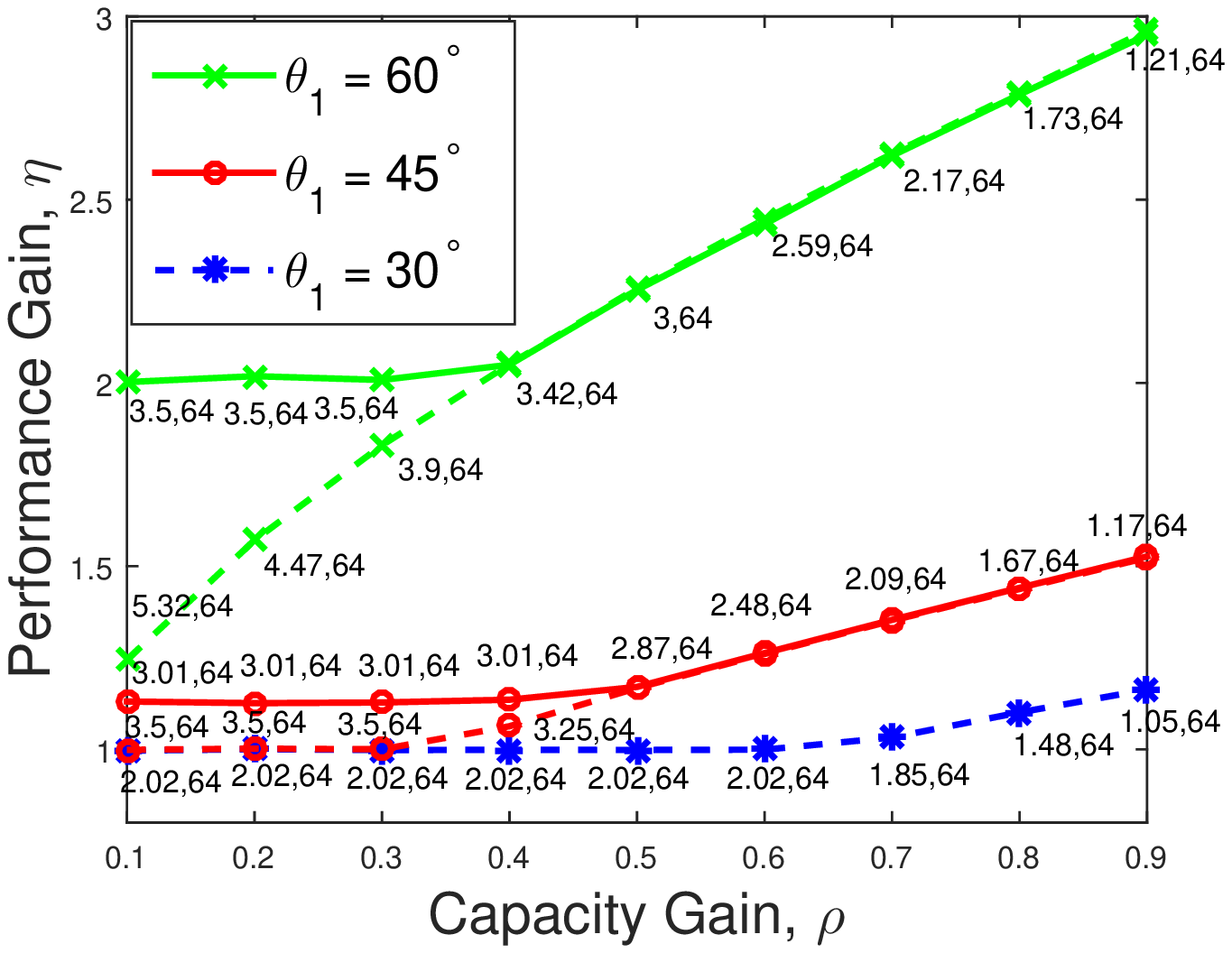}\label{AES_LOS_rho}}\\
\subfloat[No illumination constraints are imposed, $\theta_1 = 60^{\circ}$]{\includegraphics[width=0.65\textwidth]{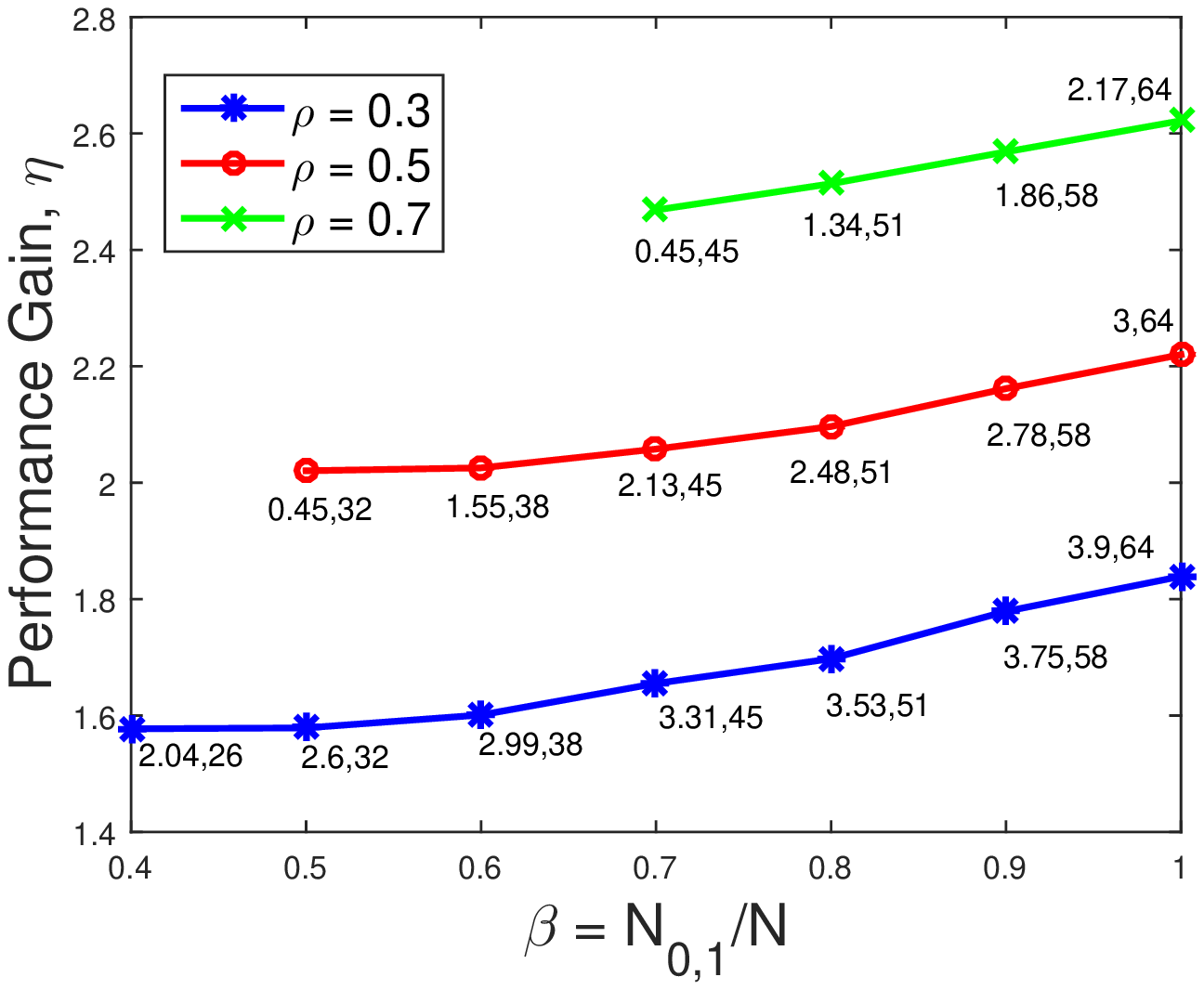}\label{AES_LOS_beta}}\\
\caption{{Performance gain considering different parameters and constraints for $N = 64.$}}
\label{fairness}
\end{figure}
Algorithm $1$ shows detailed steps to evaluate $\eta$ considering different settings and constraints.
In \figurename~\ref{AES_LOS_rho}, we illustrate $\eta$ as a function of the capacity gain (\textit{i.e.}, 
$\rho$), and  for different values of the LED viewing angle (\textit{i.e.}, $\theta_1$). We consider an equal power allocation scheme among users in both $\Z$ and $\mathcal{Z}'_1(\rho)$, and we have the results when the illumination constraints are absent (\textit{dashed lines}), and when they are present (\textit{solid lines}). For each value of $\rho$, we further show the combination $(r_{0,1},N_{0,1})$, which defines $\Z$ for the given $\rho$. \tcr{Recall that, in \figurename~\ref{AES_LOS_rho} we consider the strategy of finding the maximum possible radius $r_{0,1}$ for a given $\rho$ such that $N_{0,1} \leq N$.} Here, we assume that the interfering AP, \textit{i.e.}, AP2, is far enough to be neglected. Herein, higher values of $\eta$ implies a better performance of the proposed scheme over the benchmark scenario in terms of the area spectral efficiency, such that $\eta = 1$ means that both scenarios have the same performance. As clearly seen, the performance gain increases with $\rho$ with all values of $\theta_1$, while the gain is higher at wider angles and higher values of $\rho$. On the other hand, we observe that no performance gain can be achieved at lower values of $\rho$ with smaller $\theta_1$, \textit{i.e.}, $\eta = 1$. This can be clearly seen for $\rho \leq 0.6 $ when $\theta_1 = 30^{\circ}$ and for $\rho \leq 0.4$ when 
$\theta_1 = 45^{\circ}$. In both cases, the radius of $\Z$ equals the cell radius. These two observations agree with the initial expectation drawn from the results obtained in 
\figurename~\ref{LOS_Ns_different_rho_theta}, where the area of Zone $0$ was shown to be remarkably smaller than the cell area at smaller values of $\rho$ and/or wider LED angles. Nevertheless, the performance gain is noticeably improved at lower values of $\rho$ when the illumination constraints are present, as revealed in \figurename~\ref{AES_LOS_rho}. This is expected since the radius of $\Z$ has lower upper bounds, as expressed in \eqref{illumination_4}. Notice that the illumination constraint has no influence when $\theta_1 = 30^{\circ}$ since the upper bound is less than the cell radius.    


As shown in \figurename~\ref{AES_LOS_rho}, the high performance gain is achieved at the expense 
of the number of the subcarriers in $\mathcal{Z}'_1(\rho)$. For instance, when $\theta_1 = 60^{\circ}$, no subcarriers 
are allocated in $\mathcal{Z}'_1(\rho)$ for any value of $\rho$. However, leaving $\mathcal{Z}'_1(\rho)$ without subcarriers is not 
recommended since the subcarriers in this zone can be used for different purposes, \textit{e.g.}, to support the handover process when a user moves to another AP\cite{Marwan_ICC17}. 
Alternatively, in \figurename~\ref{AES_LOS_beta}, we assume that $\Z$ is allocated 
a predefined fraction of the subcarriers, $\beta = \frac{N_{0,1}}{N} < 1$, and we depict the performance gain 
with respect to $\beta$ for different values of $\rho$ and $\theta_1 = 60^{\circ}$. It is easily seen that the performance gain decreases with decreasing $\beta$ since less subcarriers are allocated in $\Z$, but with this approach we guarantee allocating some subcarriers at $\mathcal{Z}'_1(\rho)$. For example, we have $N_{1,1} = 6$ when $\beta = 0.9$.

\begin{figure}
\centering
\includegraphics[width=0.8\textwidth]{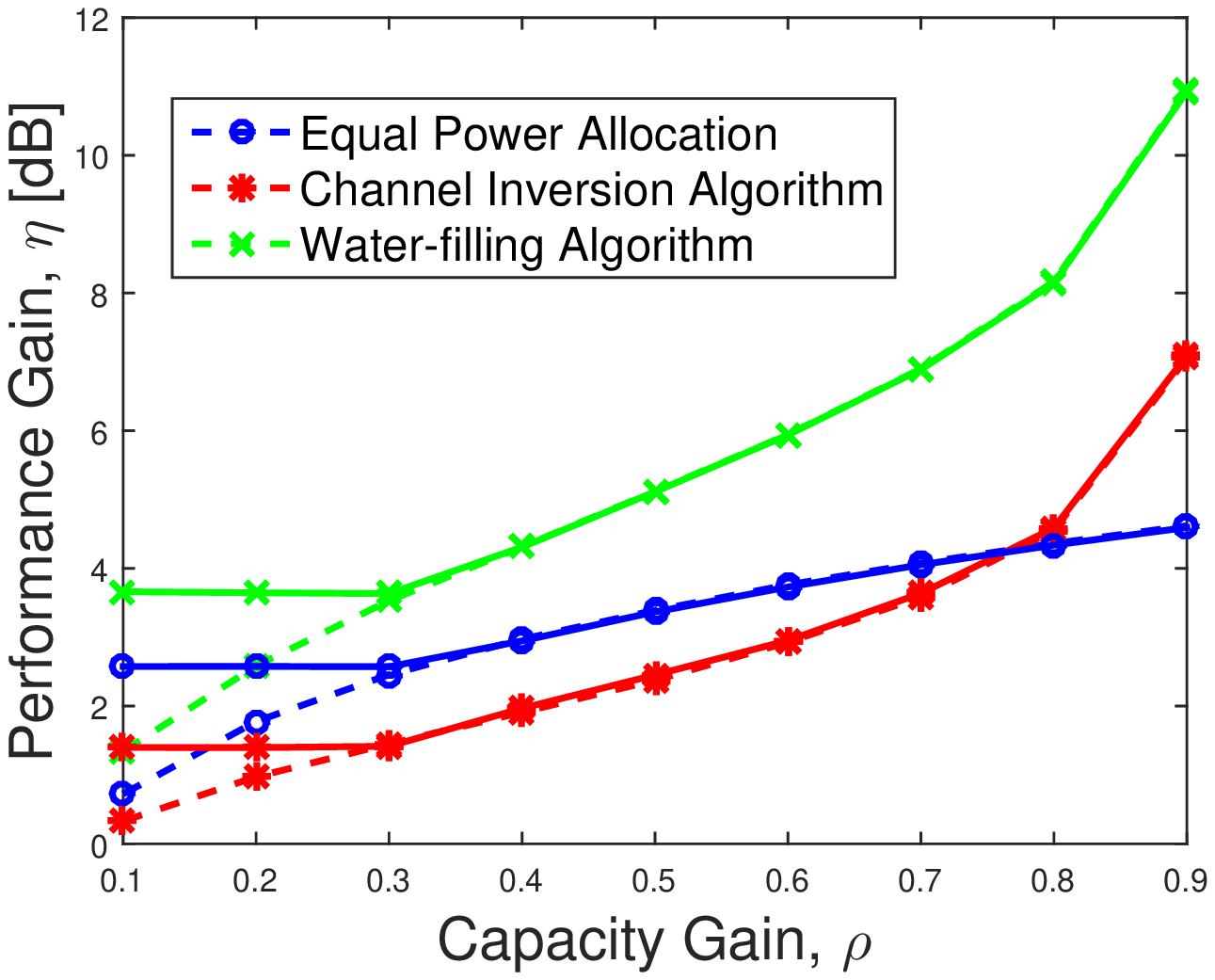}
\caption{Performance gain as a function of $\rho$ considering different power allocation schemes in $\Z$ and $\mathcal{Z}^{'}_1(\rho)$ for $N = 64$ and $\theta_1 = 60^{\circ}$. Solid (dashed) lines represent that case when the illumination constraints are present (absent).}
\label{AES_LOS_rho_PA}
\end{figure}

In \figurename~\ref{AES_LOS_rho_PA}, we illustrate the effect of applying different per-user power allocations schemes in Phase $2$ (\textit{User Level}) on the gain performance for $\theta_1 = 60^{\circ}$. In particular, we consider two standard power allocation techniques \textit{i.e.}, the \textit{water-filling algorithm}, and the \textit{channel inversion algorithm}~\cite{tse2005fundamentals}. In water-filling scheme, more power is allocated to users with higher channel gains. Noting that the channel gain is inversely proportional with the square of the distance between the AP and the user, then more power is allocated to users closer to the AP. On the other hand, users closer to the AP are allocated less power in the channel inversion scheme, since users with weaker channels (further from the AP) have higher priority 
for the sake of 
fairness~\cite{tse2005fundamentals}. We further compare the results with those obtained  when the \textit{equal power allocation scheme} is applied, as also assumed in \figurename~\ref{AES_LOS_rho} and \figurename~\ref{AES_LOS_beta}. Here, we set $\beta = 1$.  As clearly seen in 
\figurename~\ref{AES_LOS_rho_PA}, the water-filling scheme 
has the best performance for all values of $\rho$. The channel inversion scheme has the worst performance for $\rho<0.8$, whereas the performance is improved at higher values. 
\figurename~\ref{AES_LOS_rho_PA} agrees with the well-known trade off between the achieved rates and the fairness among users. Again, we observe that imposing constraints on the illumination level improves the system performance at lower values of $\rho$.  

\tcr{\subsection{Fairness Measure}}
Now, we investigate the performance of the proposed scheme from the user perspective. Noting that the proposed algorithm allocates more resources in Zone $0$ than those allocated in Zone $1$, 
especially at larger values of $\rho$, then the \textit{fairness} issues arise as a fundamental criteria that should be taken into account. To that end, we consider the ratio between the average rate 
achieved in $\mathcal{Z}_1^{'}(\rho)$ to the average rate achieved in $\Z$, denoted as $\zeta$, as the fairness measure. Notice that larger values of $\zeta$ represent higher levels of fairness supported by the proposed scheme, with $\zeta = 1$ refers tot the ideal case when both zones have the same average achievable rate.   
\begin{figure}
\centering
\subfloat[Equal Power Allocation]{\includegraphics[width=0.55\textwidth]{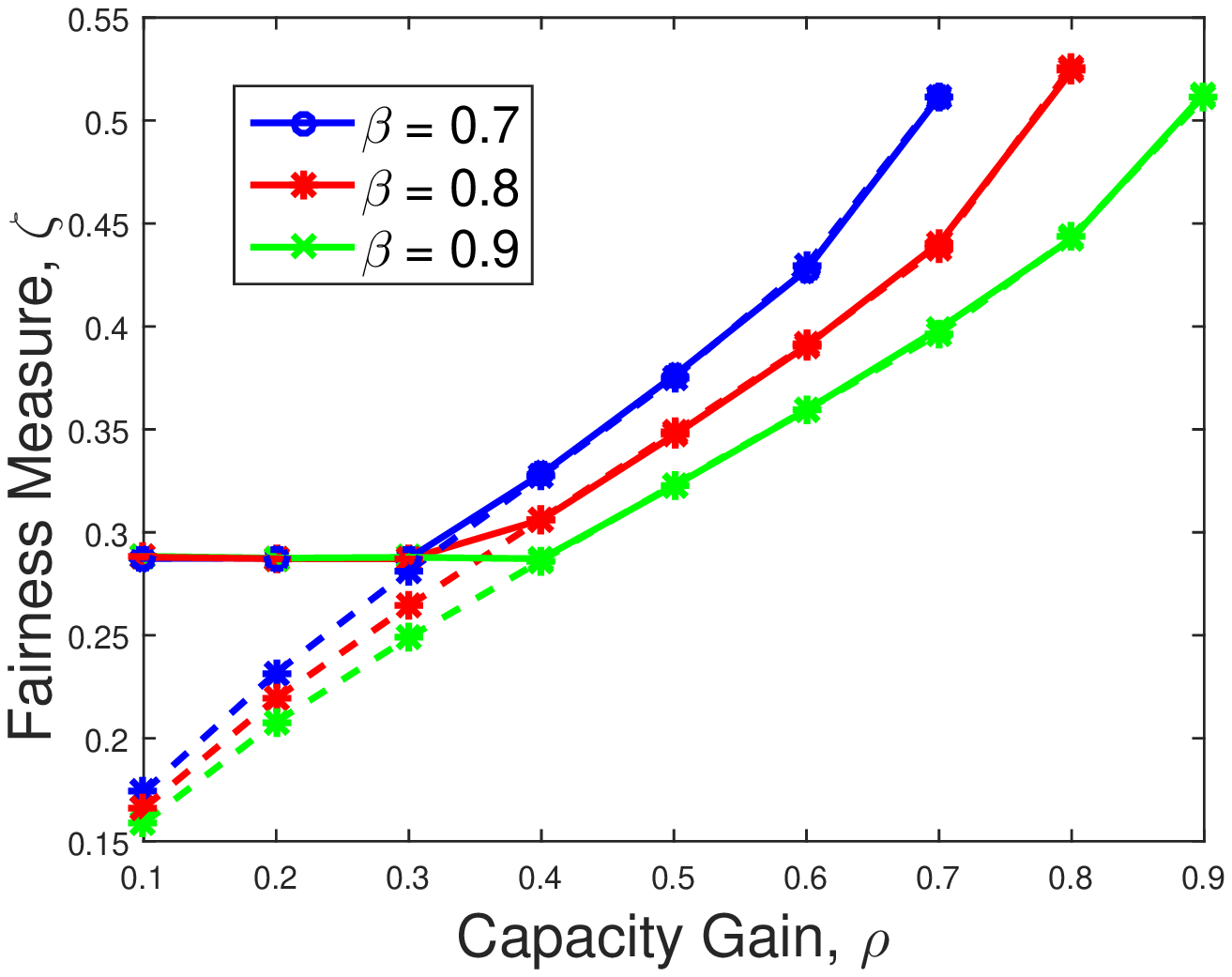}\label{fairness_eq_power}}
\subfloat[Different Power Allocation Schemes and $\beta = 0.9$]{\includegraphics[width=0.55\textwidth]{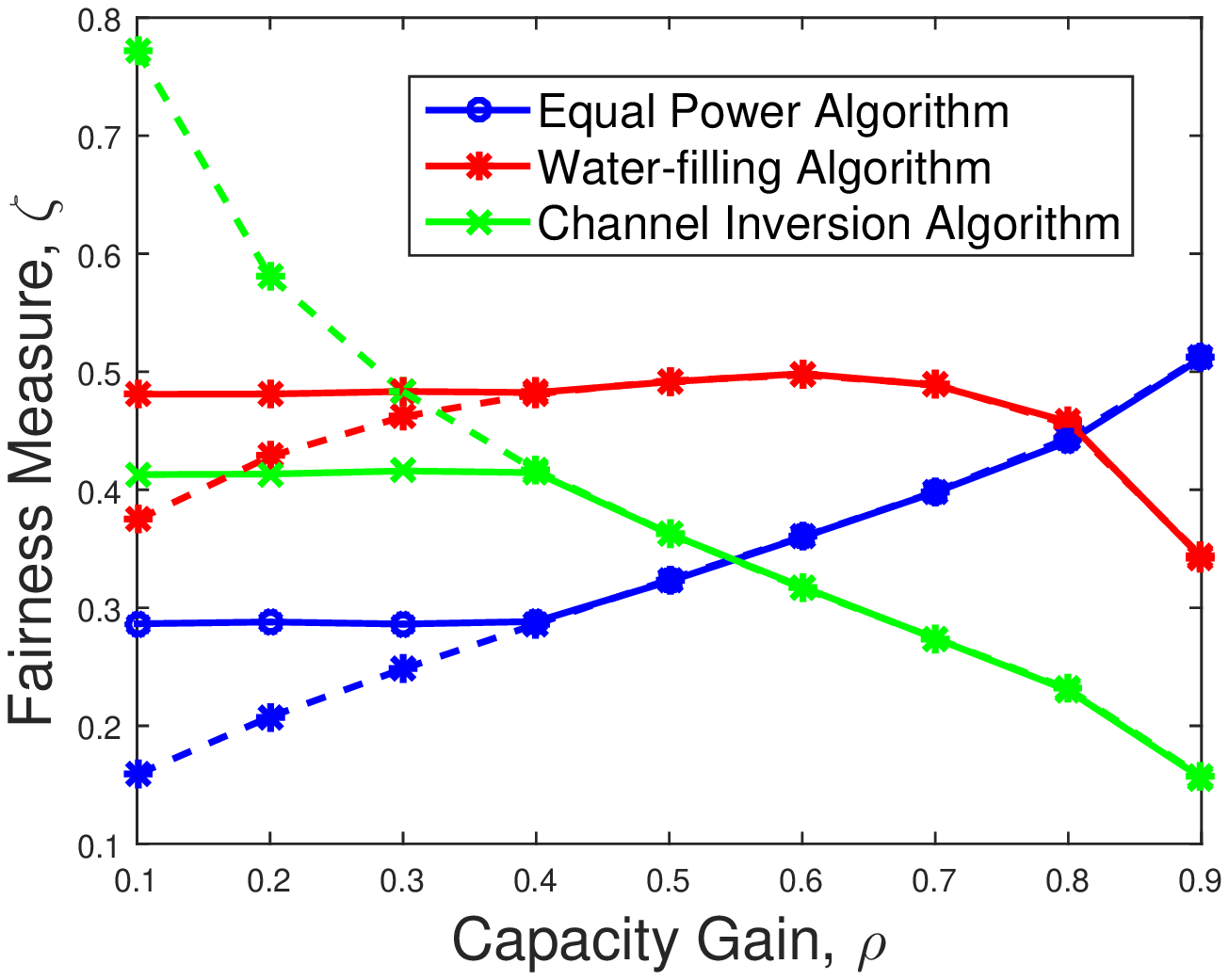}\label{fairness_PA}}
\caption{{Fairness measure $\zeta$ versus $\rho$ for $\theta_1 = 60^{\circ}$ and $N = 64$. Solid (dashed) lines represent the case of present (absent) illumination constraints.}}
\label{fairness}
\end{figure}

In \figurename~\ref{fairness}, we depict the fairness measure as a function of the capacity gain for different illumination conditions (\textit{i.e.}, absence and presence of illumination constraints), and when $\theta_1 = 60^{\circ}$. Here, we distinguish the results when an equal power allocation scheme among all users is applied with different values of $\beta$ (\figurename~\ref{fairness_eq_power}), and when different power allocation schemes are applied for $\beta = 0.9$ (\figurename~\ref{fairness_PA}). From \figurename~\ref{fairness_eq_power}, for increasing $\rho$ and when an equal power allocation scheme is applied, we observe an increase of the fairness measure  for all values of $\beta$ and different illumination conditions. This observation agrees with the fact that increasing $\rho$ results in decreasing 
the radius of $\Z$, and then more users located in Zone $1$ can get closer to the AP and hence achieve the average rate achieved in Zone $1$. \tcr{In addition, notice that reducing $\beta$ has a potential impact on improving the fairness measure, even though the majority of the resources are still allocated in Zone 0, as observed from Figure 4b.} 

On the other hand, 
in \figurename~\ref{fairness_PA} when the water-filling algorithm is applied (red line), we notice a potential enhancement in the fairness level over the equal power allocation scheme 
(blue line) for $\rho < 0.8$. However, the performance is getting worse at higher values of $\rho$, which can be explained since the area of $\mathcal{Z}_1^{'}(\rho)$ increases and, hence, more users located in Zone $1$ get close to the cell edges with less allocated power. Therefore, the average rate achieved in Zone $1$ is reduced and the fairness level decreases.  
Finally, we observe that applying the channel inversion method (green line 
in \figurename~\ref{fairness_PA}) results in the best fairness levels at lower values of $\rho$ (\textit{i.e.}, 
$\rho\leq 0.4$), whereas the performance is getting worse with increasing $\rho$. Indeed, even though more users are getting closer to the cell center,  they are allocated less power since users closer to the edge have more priority. 

We remark  that, while the channel inversion scheme aims at increasing the fairness among users in each zone, $\zeta$ represents the fairness level between both zones in terms of the per-user average rate in each zone. Therefore, increasing $\rho$ can increase the total rate achieved in both zones, as depicted in \figurename~\ref{AES_LOS_rho_PA}, but the per-user average rate decreases, as shown in \figurename~\ref{fairness_PA}. 
Through a comparison between \figurename~\ref{AES_LOS_rho_PA} and  
\figurename~\ref{fairness}, we can  observe that, increasing $\rho$ can improve both  the ASE and the fairness level between $\Z$ and $\mathcal{Z}_1^{'}(\rho)$ in most of the scenarios. However, increasing $\rho$ results in smaller areas of $\Z$, as shown in \figurename~\ref{AES_LOS_rho}, and hence allocating the majority of the resources (subcarriers and power) in smaller geographical areas. This strategy might not be practically feasible, especially with small values of $\theta_1$. Subsequently, obtaining the optimal values of $\rho$ and $\beta$ that maximize that ASE, while achieving a sufficient level of fairness in practically feasible scenarios, is a key requirement for the proposed scheme.

\begin{figure}[t]
\centering
\includegraphics[width=0.8\columnwidth]{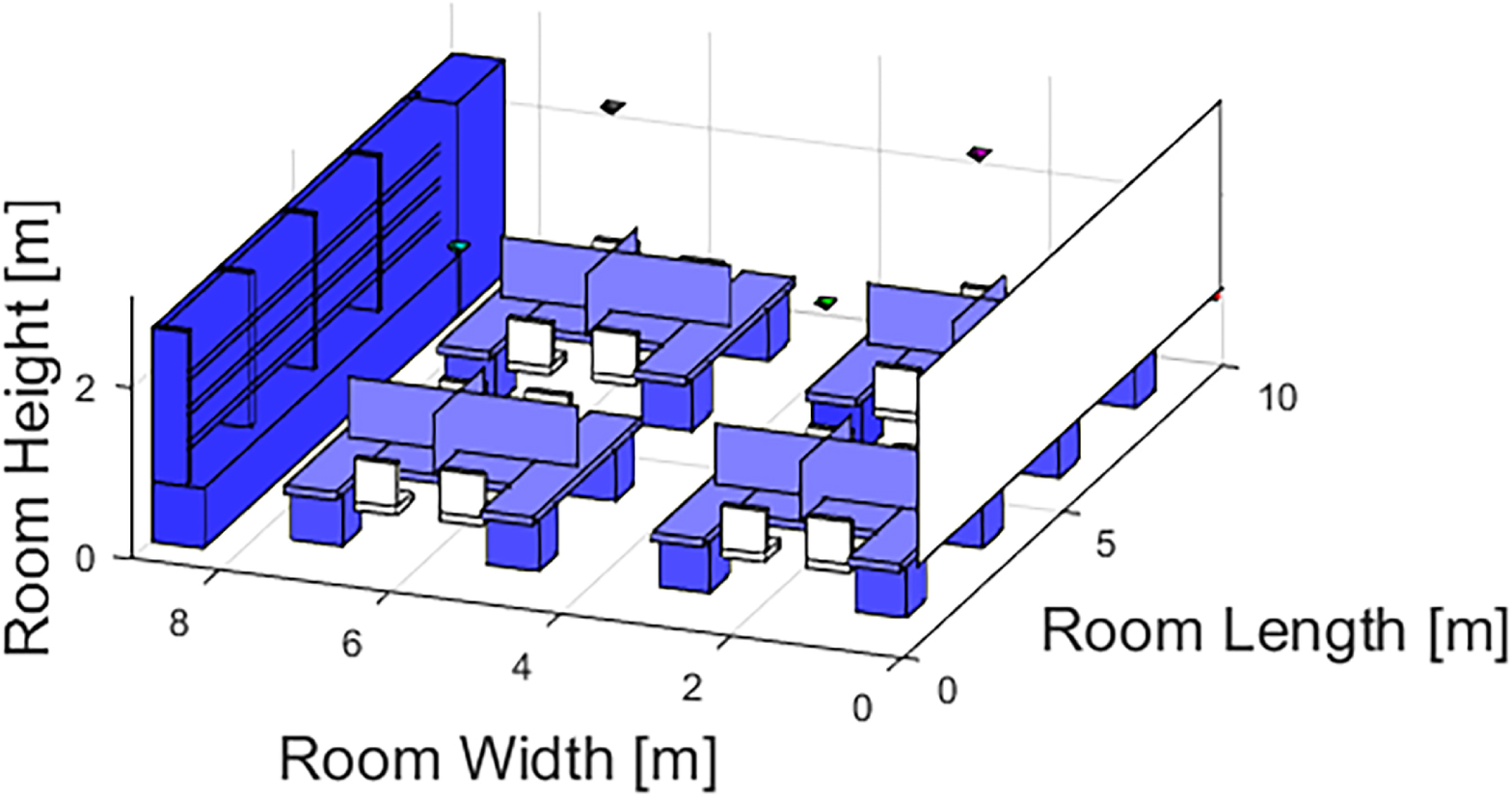}
\caption{Simulation scenario for a furnished indoor office from the Department of Engineering at Roma Tre University.}
\label{System_Candles}
\end{figure}

\begin{figure}
\centering
\subfloat[LoS only, $\rho = 0.1$]{\includegraphics[width=0.4\columnwidth]{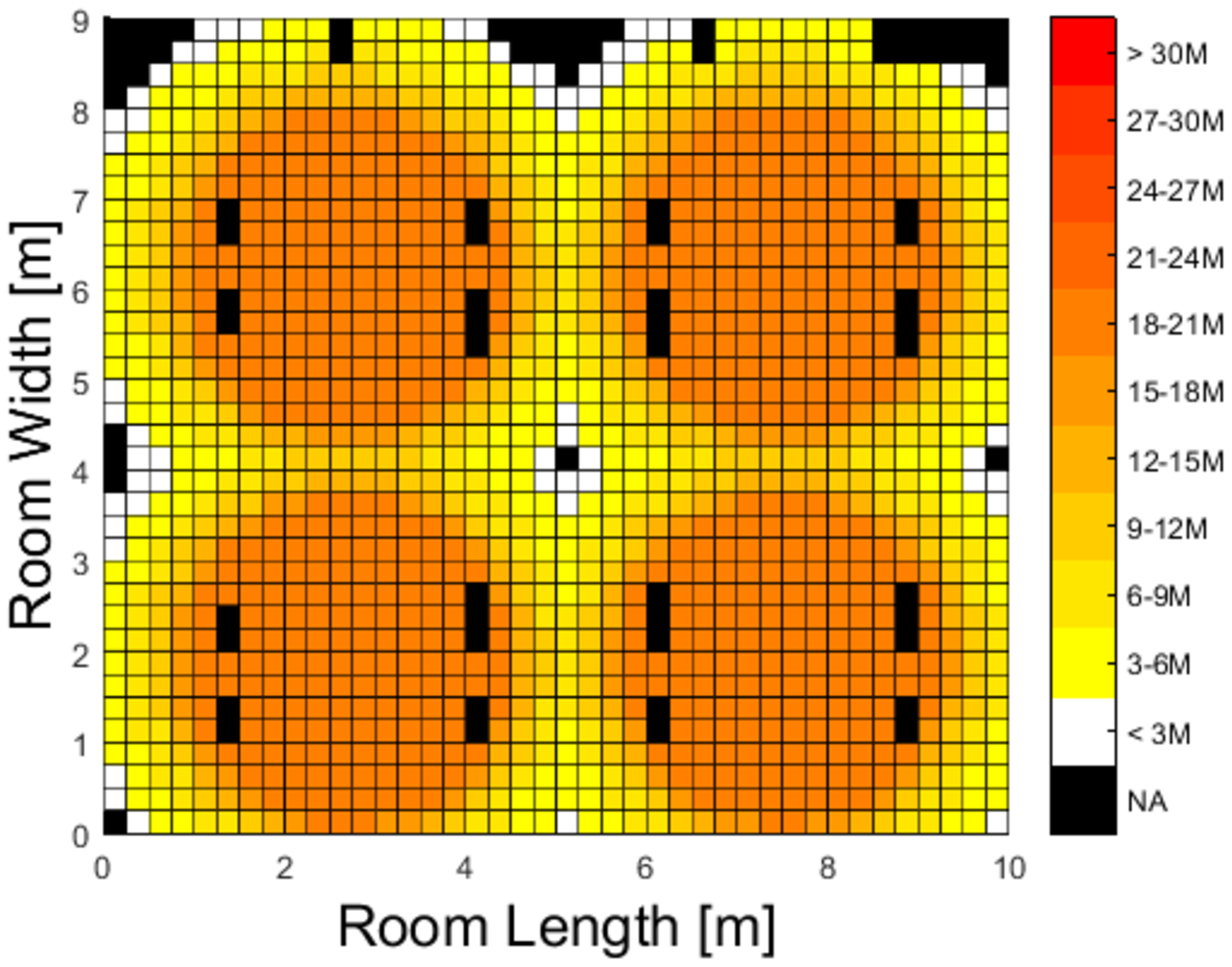}}
\subfloat[LoS only, $\rho = 0.5$]{\includegraphics[width=0.4\columnwidth]{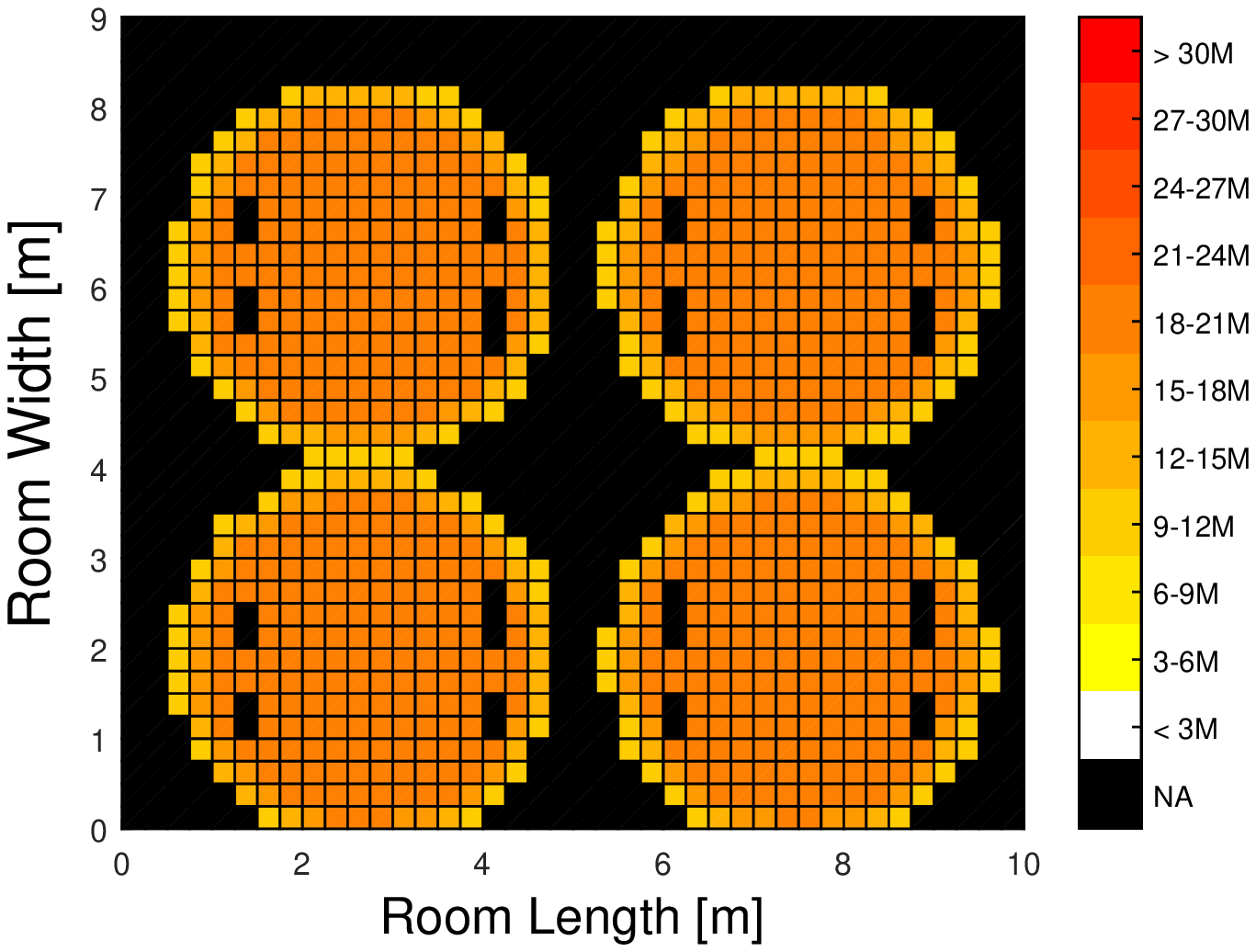}}
\subfloat[LoS only, $\rho = 0.9$]{\includegraphics[width=0.4\columnwidth]{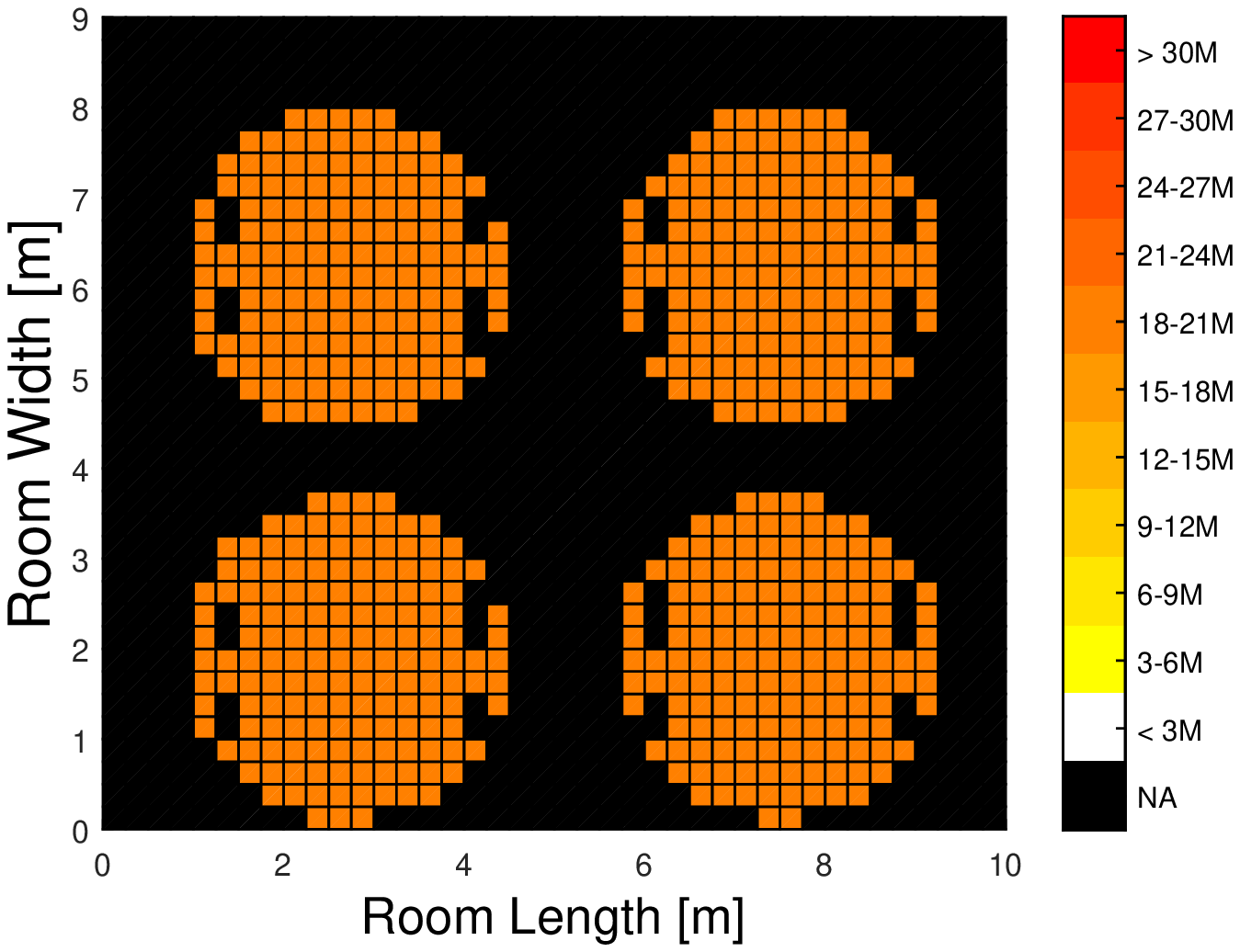}}\\
\subfloat[LoS and Diffuse, $\rho = 0.1$]{\includegraphics[width=0.4\columnwidth]{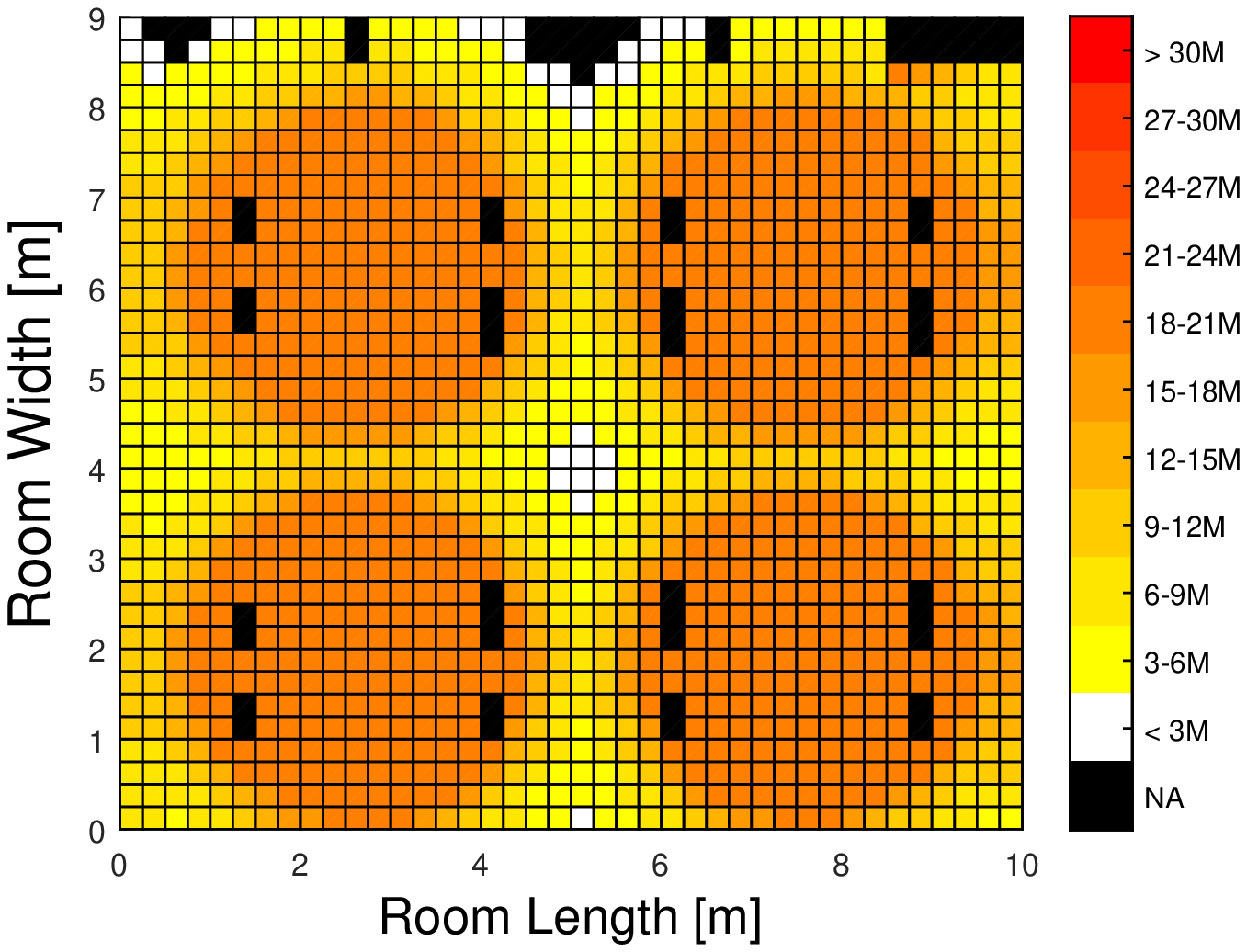}} 
\subfloat[LoS and Diffuse, $\rho = 0.5$]{\includegraphics[width=0.4\columnwidth]{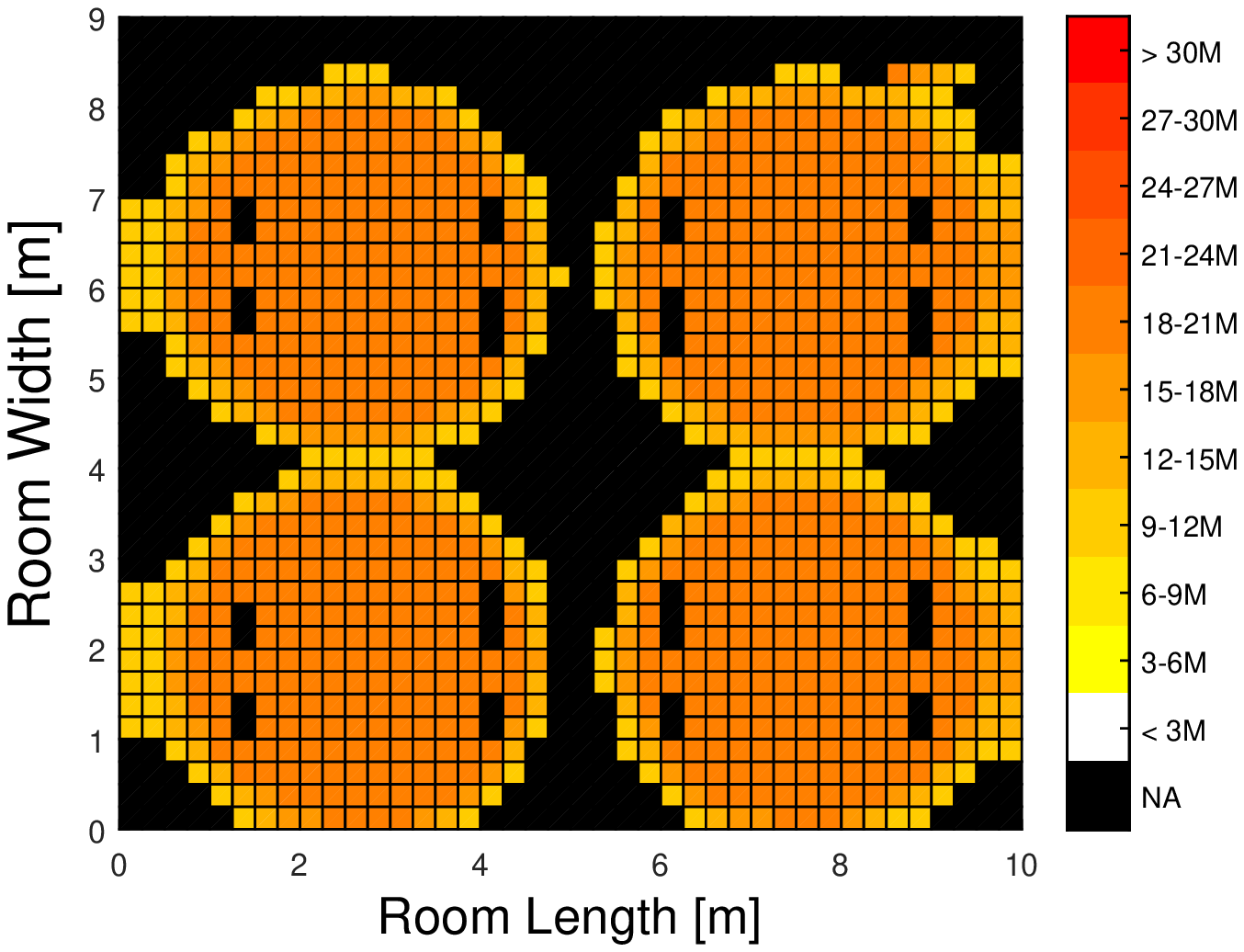}}
\subfloat[LoS and Diffuse, $\rho = 0.9$]{\includegraphics[width=0.4\columnwidth]{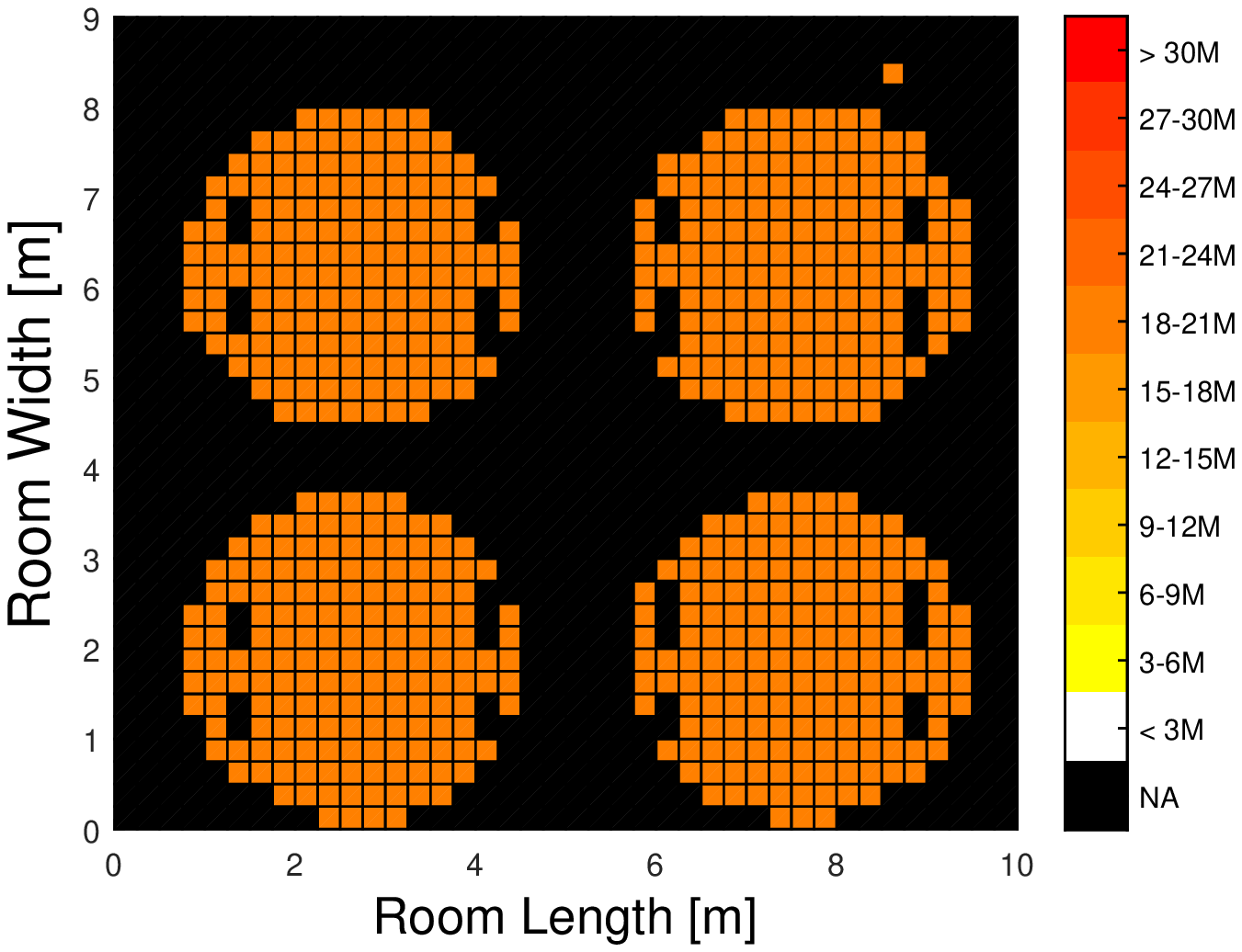}} 
\caption{Data rate distribution in the simulated indoor scenario, considering different values of $\rho$ and 
different propagation modes.}
\label{Zone_0_Candles}
\end{figure}

\subsection{Realistic simulation scenario}
\label{sec:Los_nlos}
Finally, in this subsection we investigate the system performance when applying the proposed scheme on a realistic indoor scenario.  
The reference scenario is a $10\times9\times3$~m$^3$ furnished office room with four LED-based APs installed at the ceiling with locations $(2.7,1.9,3)$, 
$(2.7,6.2,3)$, $(7.5,1.9,3)$, and $(7.5,6.2,3)$~m, 
as shown in 
\figurename~\ref{System_Candles}. The scenario is modeled through the communication and lighting emulation 
software (CandLES)~\cite{rahaim2010candles}. All APs have the viewing 
angle of $\theta = 60^{\circ}$ and they all employ the ON-OFF keying (OOK) modulation schemes. 
Transmitting and receiving planes are parallel to each other and the distance between them  is $2$~m.

In \figurename~\ref{Zone_0_Candles}, we initially display 
the data rate distribution over the entire space and we show the zones $\{\mathcal{Z}_k(\rho)\}_{k=1}^4$ for different values of $\rho$, and different 
propagation modes (\textit{i.e.},  LoS only, and LoS plus diffuse). Here, we set $N = 64$ and $\beta = 1$ for all cells. 
For diffuse links, we assume that light rays are bounced once at most before reaching the reviving plane. As expected, increasing $\rho$ results in decreasing the coverage area of the AP. In other words, 
APs provide a spotlighting coverage, which is preferred in work places like study rooms and offices. We further observe the minor effects of the diffuse transmission links, especially at the edges close to the walls.

\begin{figure}[t]
\centering
\subfloat[LoS only]{\includegraphics[width=0.55\textwidth]{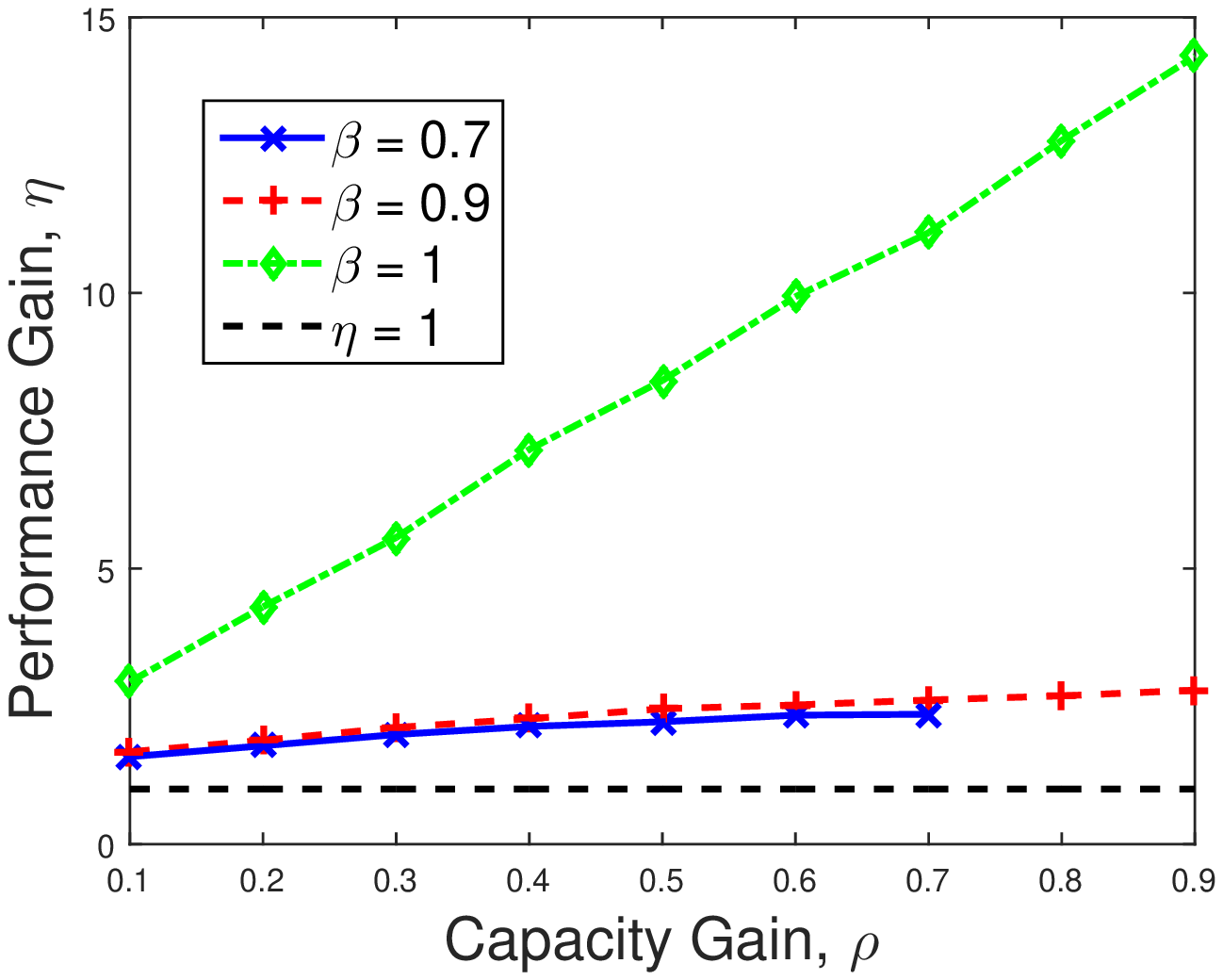}\label{AES_candles_beta_Bounce_0}}
\subfloat[LoS and Diffuse]{\includegraphics[width=0.55\textwidth]{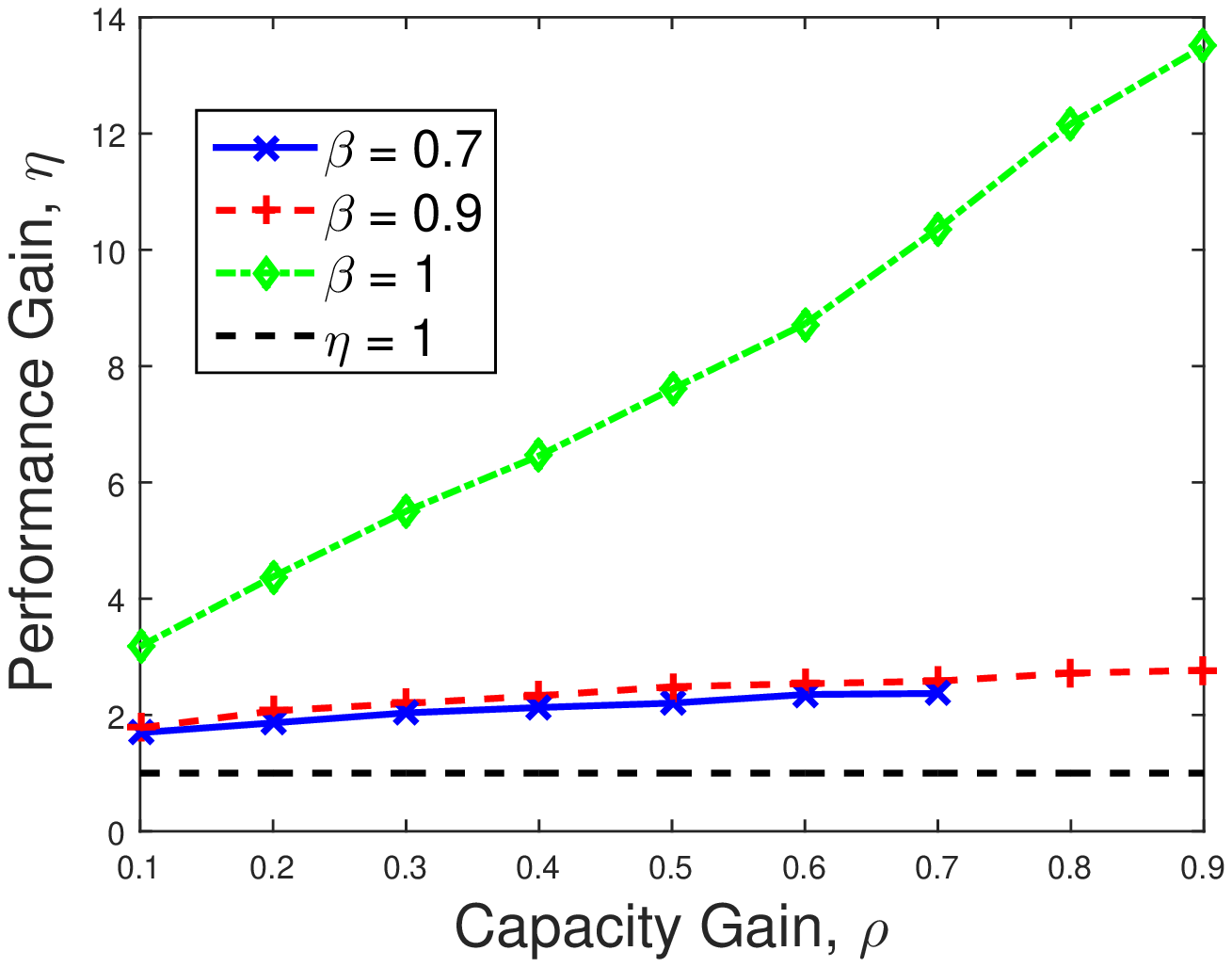}\label{AES_candles_beta_Bounce_1}}
\caption{Area spectral efficiency gain of the proposed scheme in the simulated indoor scenario as a function of $\rho$ for different values of $\beta$ and different transmission modes.}
\label{AES_candles_beta}
\end{figure}

In \figurename~\ref{AES_candles_beta}, we simulate the performance gain $\eta$ as a function of the capacity gain $\rho$ for different values 
of $\beta$. Here, we consider the equal power allocation scheme, and we show the results for the AP located at the upper-right corner in \figurename~\ref{System_Candles}, \textit{i.e.,} with location $(7.5,1.9,3)$.  We can immediately observe the high gain achieved be applying the proposed 
scheme over the reference one for all values of $\rho$. We further notice that the performance gain is slightly higher when only LoS links are considered, as well depicted in 
\figurename~\ref{AES_candles_beta_Bounce_0}, \tcr{which also confirms the initial assumption that the majority of the collected energy at the photodiode comes from the LoS component.}

\section{Conclusions}
In this paper, we have presented a resource allocation technique which is also capable of suppressing the inter-cell interference based on a certain geometrical design of the cell area. 
By dividing the coverage area into two location-based priority zones (\textit{i.e.}, Zone $0$ and Zone $1$), we have provided a two-step RA scheme in which each step corresponds to a different level 
of allocating bandwidth and power resources, \textit{i.e.}, cell-level and user-level allocation steps. We have showed detailed analyses towards defining both zones in terms of the physical area and the amount of allocated subcarriers. We have further analyzed the influence of illumination needs on defining both zones. Through simulations, we have evaluated the performance of the proposed scheme in terms of the area spectral efficiency and the fairness level between both zones, while considering different power allocation schemes among users in each zone. We have shown that we can potentially enhance both performance measures by carefully setting a certain design parameter (\textit{i.e.}, $\rho$),  which reflects the priority level 
of the users located in Zone $0$, in terms of the achievable rates. Finally, we have applied the proposed scheme in a realistic indoor scenario and evaluated the performance using a simulation tool. 
As a future work, we highlight the following two main directions: (\textit{i}) conducting a more detailed analytical study towards optimizing the system parameters (\textit{i.e.,} $\rho$, $r_{0,k}$, and $N_{0,k}$) and performance, and \textit{ii)} providing a proof-of-concept experimental study validating the effectiveness of the proposed scheme in real-time scenarios.

\section*{References}


\bibliography{references}

\end{document}